\documentclass{aa}
\AtBeginDocument{\nolinenumbers\let\linenumbers\relax}
\usepackage[utf8]{inputenc}
\usepackage{listings}
\usepackage{xcolor}
\usepackage{stfloats}
\usepackage{graphicx}
\usepackage{txfonts}
\usepackage[breaklinks=true,colorlinks=true,linkcolor=blue,citecolor=magenta,urlcolor=blue,pdfborder={0 0 0}]{hyperref}

\lstdefinestyle{adql}{
    basicstyle=\ttfamily\scriptsize,
    keywordstyle=\color{blue}\bfseries,
    breaklines=true,
    breakatwhitespace=true,
    columns=fullflexible,
    frame=single,
    captionpos=b,
    keywords={SELECT,AS,FROM,INNER,LEFT,JOIN,ON,WHERE,AND,OR}
}
\usepackage{placeins}
\newif\ifnotes
\notestrue

\usepackage{orcidlink}
\usepackage{float}
\newcommand{\jplus}{\mbox{J-PLUS}}
\newcommand{\splus}{\mbox{S-PLUS}}
\newcommand{\gaia}{\textit{Gaia}}
\newcommand{\teff}{\ensuremath{T_{\mathrm{eff}}}}
\newcommand{\logg}{\ensuremath{\log g}}
\newcommand{\feh}{\ensuremath{[\mathrm{Fe/H}]}}
\newcommand{\am}{\ensuremath{[\alpha/\mathrm{M}]}}
\newcommand{\afe}{\ensuremath{[\alpha/\mathrm{Fe}]}}

\newcommand{\smad}{\ensuremath{\sigma_{\rm MAD}}}
\newcommand{\snr}{\ensuremath{\mathrm{S/N}}}
\renewcommand{\topfraction}{0.9}
\renewcommand{\bottomfraction}{0.8}
\renewcommand{\textfraction}{0.07}
\renewcommand{\floatpagefraction}{0.75}
\begin{document}

   \title{\jplus: Photometric stellar parameters for nearly 15 million stars in the fourth data release}
\authorrunning{A. Hidalgo-Pinilla et al.}

\author{A. ~Hidalgo-Pinilla\inst{\ref{CEFCA}}\corrauth{ahidalgo@cefca.es}
    \and A. ~del~Pino\inst{\ref{IAA}}\email{apino@iaa.csic.es}
    \and J. ~A.~Alz\'ate-Trujillo\inst{\ref{IAA}}\email{andresalzate3146@proton.me}
    \and C. ~L\'opez-Sanjuan\inst{\ref{CEFCA},\ref{UA}}\email{clsj@cefca.es}
    \and A. ~M.~Mart\'inez-Garc\'ia\inst{\ref{IAA}}\email{ammartinez@iaa.csic.es}
    \and F. ~Arizo\inst{\ref{CEFCA}}\email{farizo@cefca.es}
    \and A. ~Hern\'an-Caballero\inst{\ref{CEFCA},\ref{UA}}\email{ahernan@cefca.es}
    \and E. ~Machado-Pereira\inst{\ref{ED}}\email{eduardompereira93@gmail.com}  
    \and S. ~Zarattini\inst{\ref{CEFCA},\ref{UA}}\email{szarattini@cefca.es}
    \and J. ~F.~Aguilar\inst{\ref{UAM},\ref{UMNG}}\email{john.aguilar@unimilitar.edu.co}
    \and A.~Alvarez-Candal \inst{\ref{IAA}}\email{varobes@gmail.com}  
    \and B. ~Anguiano \inst{\ref{CEFCA},\ref{UA}}\email{banguiano@cefca.es}
    \and V. ~M.\ Placco\inst{\ref{VP}}\email{vmplacco@gmail.com} 
    \and P. ~Cruz\inst{\ref{P}}\email{pcruz.astro@gmail.com}
    \and A.~J.~Cenarro\inst{\ref{CEFCA},\ref{UA}}
    \and D.~Crist\'obal-Hornillos\inst{\ref{CEFCA}}
    \and R.~A.~Dupke\inst{\ref{ON},\ref{MU}}
    \and A.~Ederoclite\inst{\ref{CEFCA}}
    \and C.~Hern\'andez-Monteagudo\inst{\ref{IAC},\ref{ULL}}
    \and A.~Mar\'{\i}n-Franch\inst{\ref{CEFCA},\ref{UA}}
    \and M.~Moles\inst{\ref{CEFCA}}
    \and J.~Varela\inst{\ref{CEFCA}}
    \and H.~V\'azquez~Rami\'o\inst{\ref{CEFCA},\ref{UA}}
    \and J.~Zaragoza-Cardiel\inst{\ref{CEFCA},\ref{UA}}
    \and J.~Alcaniz\inst{\ref{ON}}
    \and L.~Sodr\'e~Jr.\inst{\ref{USP}}
    \and R.~E.~Angulo\inst{\ref{DIPC},\ref{ikerbasque}}}

\institute{Centro de Estudios de F\'{\i}sica del Cosmos de Arag\'on (CEFCA), Plaza San Juan 1, 44001 Teruel, Spain\label{CEFCA}
	\and Instituto de Astrof\'isica de Andaluc\'ia (IAA-CSIC), Glorieta de la Astronom\'ia s/n, 18008 Granada, Spain\label{IAA}
	\and Unidad Asociada CEFCA-IAA, CEFCA, Unidad Asociada al CSIC por el IAA y el IFCA, Plaza San Juan 1, 44001 Teruel, Spain\label{UA}
	\and Instituto Nacional de Pesquisas Espaciais, Av. dos Astronautas 1758, Jardim da Granja,12227-010 São José dos Campos, SP, Brazil\label{ED}
	\and Universidad Aut\'onoma de Madrid, Ciudad Universitaria de Cantoblanco, 28049 Madrid, Spain\label{UAM}
	\and Departamento de Matem\'aticas, Universidad Militar Nueva Granada, kil\'ometro 2 v\'ia Cajic\'a--Zipaquir\'a, 110111, Colombia\label{UMNG}
	\and NSF NOIRLab, Tucson, AZ 85719, USA \label{VP}
	\and Telespazio UK S.L. for European Space Agency (ESA), Camino bajo del Castillo, s/n, Urbanización Villafranca del Castillo, Villanueva de la Cañada, 28692 Madrid, Spain \label{P}
	\and Observat\'orio Nacional - MCTI (ON), Rua Gal. Jos\'e Cristino 77, S\~ao Crist\'ov\~ao, 20921-400 Rio de Janeiro, Brazil\label{ON}
	\and University of Michigan, Department of Astronomy, 1085 South University Ave., Ann Arbor, MI 48109, USA\label{MU}
	\and Instituto de Astrof\'{\i}sica de Canarias, La Laguna, 38205, Tenerife, Spain\label{IAC}
	\and Departamento de Astrof\'{\i}sica, Universidad de La Laguna, 38206, Tenerife, Spain\label{ULL}
	\and Instituto de Astronomia, Geof\'{\i}sica e Ci\^encias Atmosf\'ericas, Universidade de S\~ao Paulo, 05508-090 S\~ao Paulo, Brazil\label{USP}
	\and Donostia International Physics Centre (DIPC), Paseo Manuel de Lardizabal 4, 20018 Donostia-San Sebasti\'an, Spain\label{DIPC}
	\and IKERBASQUE, Basque Foundation for Science, 48013, Bilbao, Spain\label{ikerbasque}
}


\abstract
{}
{We aim to derive the effective temperature (\teff), surface gravity (\logg), metallicity (\feh), and $\alpha$-element abundance (\am), with uncertainties and correlations, for $\sim$14.6 million stars with $r \leq 19$~mag in the fourth data release (DR4) of the Javalambre Photometric Local Universe Survey (\jplus).}
{We fitted degree-4 polynomials in subsets of the 11 \jplus\ colours relative to the $r$ band, and of the 11 principal components (PCs) of the 66 possible colours, to 33,926 stars with parameters from the Large Sky Area Multi-Object Fiber Spectroscopic Telescope (LAMOST) medium-resolution survey DR12, selecting for each parameter the subset producing the best results. The uncertainties come from a Monte Carlo propagation of the photometric, calibration, extinction, and coefficient errors, calibrated against the spectroscopic residuals.}
{On an independent sample of 8,031 LAMOST stars, the colour models recover \teff, \logg, \feh, and \am\ with $R^2 = 0.969$, 0.915, 0.895, and 0.697 and a robust scatter of 73~K, 0.11~dex, 0.067~dex, and 0.025~dex; the PC models perform equivalently. The predictions agree with Apache Point Observatory Galactic Evolution Experiment DR17, GALactic Archaeology with HERMES DR4, and independent results with J-PLUS DR3 with median offsets below 60~K and 0.06~dex. Quality flags define a recommended sample of nearly 5 million sources that reproduces the known structure of the Kiel diagram and the \feh--\am\ plane and allows a chemical separation of the thin disc, the thick disc, and the halo down to $r \approx 17.5$~mag.}
{The catalogue provides the four parameters, their uncertainties, correlations, and quality flags for the 14,628,650 \jplus\ DR4 sources with $r \leq 19$~mag. The polynomial coefficients and covariance matrices are also released, so that the parameters can be recomputed and the models applied to other surveys with the same filter system.}


\maketitle

\section{Introduction}

The Milky Way grew by accreting smaller galaxies \citep{2018MNRAS.478..611B,2018Natur.563...85H}, and the record of that growth is preserved in its stars: their atmospheres retain the chemical composition of the gas from which they formed, and their orbits keep memory of where they were born. Reading this record has already reshaped our picture of the Galaxy. More than 80\,\% of the stellar halo beyond 15~kpc from the Galactic centre is now attributed to just two accreted dwarf galaxies \citep{2020ApJ...901...48N}, and the bimodal distribution of the disc stars in the plane of metallicity, \feh, and $\alpha$-element abundance, \am, encodes the different enrichment histories of the thick and thin discs \citep{2015ApJ...808..132H,2022Natur.603..599X,2025A&A...704A.258F,Alzate-trujillo2026}. Separating these populations requires for each star both its kinematics and a minimal set of physical and chemical parameters: the effective temperature, \teff, the surface gravity, \logg, \feh, and \am. The last one is especially valuable, because it distinguishes populations enriched mainly by core-collapse supernovae from those that also received the products of Type~Ia supernovae, and therefore separates accreted from in-situ stars, and thick-disc from thin-disc stars, at the same metallicity \citep{1979ApJ...229.1046T,1986A&A...154..279M,2010A&A...511L..10N,2014A&A...572A..33M}.

The kinematic half of this information is now available for nearly 1.5 billion stars thanks to the \gaia\ mission \citep{2016A&A...595A...1G,GaiaDR3}. The chemical half is the bottleneck. Spectroscopic surveys such as the Large Sky Area Multi-Object Fiber Spectroscopic Telescope \citep[LAMOST;][]{2012RAA....12.1197C,2020arXiv200507210L}, the Apache Point Observatory Galactic Evolution Experiment \citep[APOGEE;][]{2017AJ....154...94M}, the RAdial Velocity Experiment \citep[RAVE;][]{2020AJ....160...82S}, and the GALactic Archaeology with HERMES survey \citep[GALAH;][]{2015MNRAS.449.2604D} have delivered stellar parameters and abundances for millions of stars, but they reach only a small fraction of the \gaia\ sources, for three reasons. They are shallow: most of them target stars brighter than 14th magnitude. They are fibre-limited, so each pointing yields a fixed number of spectra regardless of the stellar density. And their targets are pre-selected in colour and magnitude, which produces selection functions that are difficult to model. Ongoing and upcoming surveys such as the Dark Energy Spectroscopic Instrument \citep[DESI;][]{Cooper2023}, the WHT Enhanced Area Velocity Explorer \citep[WEAVE;][]{Jin2024}, and the 4-metre Multi-Object Spectroscopic Telescope \citep[4MOST;][]{deJong2019} will go deeper and wider, but they remain subject to the last two limitations.

Photometric surveys do not suffer from these limitations. An imaging survey observes every source in its footprint down to its flux limit, typically several magnitudes fainter than any spectroscopic survey, so its selection function reduces to that limit and the surveyed area. The price is a much coarser sampling of the spectrum. Narrow- and medium-band systems recover part of the lost information by placing filters on specific spectral features. Pristine \citep{2017MNRAS.471.2587S}, SkyMapper \citep{2007PASA...24....1K}, the Mapping the Ancient Galaxy in CaHK survey \citep[MAGIC;][]{2026arXiv260526581C}, and the Stellar Abundance and Galactic Evolution Survey \citep[SAGES;][]{2023ApJS..268....9F} follow this strategy with one or two filters centred on the metallicity-sensitive Ca\,H\,\&\,K region, and were designed primarily to find very metal-poor stars.

The Javalambre Photometric Local Universe Survey (\jplus\footnote{\url{https://www.j-plus.es}}; \citealt{Cenarro19,LopezSanJuan24}) goes a step further. Its 12-band system combines five broad bands ($u$, $g$, $r$, $i$, and $z$) with seven medium and narrow bands placed on the [\ion{O}{ii}] lines (J0378), the Ca\,H\,\&\,K lines (J0395), H$\delta$ (J0410), the carbon G band (J0430), the Mg\,b triplet (J0515), H$\alpha$ (J0660), and the Ca triplet (J0861). Together, these bands are sensitive not only to metallicity but also to temperature, surface gravity, and the abundance of $\alpha$ elements, so that the four parameters listed above can be estimated from the photometry alone. \jplus\ therefore provides a flux-limited, homogeneously selected stellar sample that extends several magnitudes deeper than the most spectroscopic surveys, which is precisely the regime where the halo and the outer disc are sampled.

Stellar parameters have already been derived with different methods from the photometry of \jplus\ \citep{2019A&A...622A.182W,2022A&A...657A..35G,2022A&A...659A.181Y,2022A&A...664A..38W,2024ApJ...974..192H,2026A&A...714A.332A,10.1093/mnras/stag132} and of its southern counterpart, the Southern Photometric Local Universe Survey \citep[\splus;][]{2019MNRAS.489..241M,2021ApJ...912..147W,2025A&A...693A.306F,Huang_2026}.

In this work we present a simple, fast, and interpretable method to derive \teff, \logg, \feh, and \am, together with their uncertainties and mutual correlations, for more than 15 million \jplus\ DR4 stars. For each parameter, the method selects the subset of \jplus\ colours, or of their principal components (PCs), that carries the relevant information, and fits a polynomial to it using stars with parameters from the LAMOST medium-resolution survey. Our work differs from the previous ones based in J-PLUS data in three respects. First, it is based on DR4, the widest \jplus\ release to date. Second, the models are explicit polynomials whose coefficients and covariance matrices are published, so that the parameters can be recomputed, and their uncertainties propagated. Third, the predictions require only the \jplus\ photometry and an extinction correction, and they include \am.

The paper is organised as follows. Section~\ref{datos} describes the \jplus\ sample and the construction of the training and test samples from LAMOST. Section~\ref{sec:models} presents the polynomial models and the estimation of the uncertainties. Section~\ref{sec:results} validates the predictions and their uncertainties against an independent test sample, Sect.~\ref{sec:catalogue} describes the catalogue and its quality flags, and Sect.~\ref{sec:comparison} compares the predictions with APOGEE, GALAH, and \citet{2024ApJ...974..192H}. Section~\ref{sec:applications} tests the catalogue at the population level, and Sect.~\ref{sec:conclusions} summarises our conclusions.

\section{Data}\label{datos}

This work rests on two data sets: the \jplus\ DR4 photometry, for which we derive stellar parameters, and the LAMOST DR12 spectroscopy, which provides the parameters used to train and test the models.

\subsection{J-PLUS DR4}\label{sec:jplus}

\jplus\ is carried out at the Observatorio Astrof\'isico de Javalambre \citep[OAJ, Teruel, Spain;][]{Cenarro14} with the 83\,cm Javalambre Auxiliary Survey Telescope (JAST80) and T80Cam, a panoramic camera of $9.2{\rm k} \times 9.2{\rm k}$ pixels that provides a 2~deg$^2$ field of view with a pixel scale of 0.55~arcsec~pix$^{-1}$ \citep{MarinFranch15}. Its fourth data release (DR4)\footnote{\url{https://www.j-plus.es/datareleases/data_release_dr4}} covers $\sim$5000~deg$^2$ of the northern sky, with median $5\sigma$ depths between 20.55 and 21.94~mag (AB), depending on the band \citep{2026arXiv260706662H}. The photometry was extracted in dual mode, using the $r$ band for detection.

\begin{figure}[t]
    \centering
    \includegraphics[width=\columnwidth]{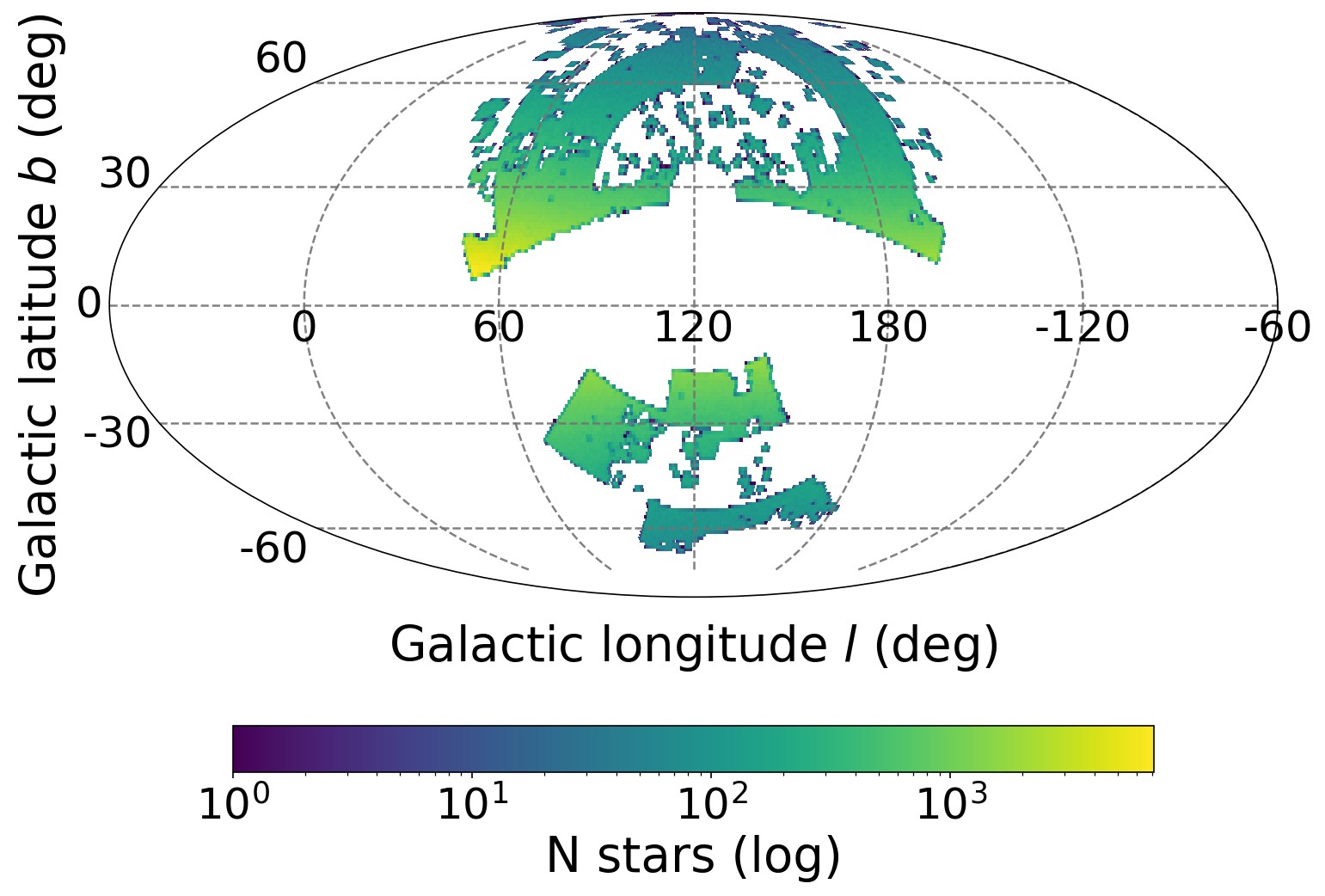}
    \caption{Distribution of the complete sample (14,628,650 stars) in Galactic coordinates, centred on $l = 120^{\circ}$, in a Mollweide projection. The colour shows the number of stars per cell on a logarithmic scale.}
    \label{fig:JPLUS_DR4}
\end{figure}

\begin{figure}[!htbp]
	\centering
	\includegraphics[width=\columnwidth]{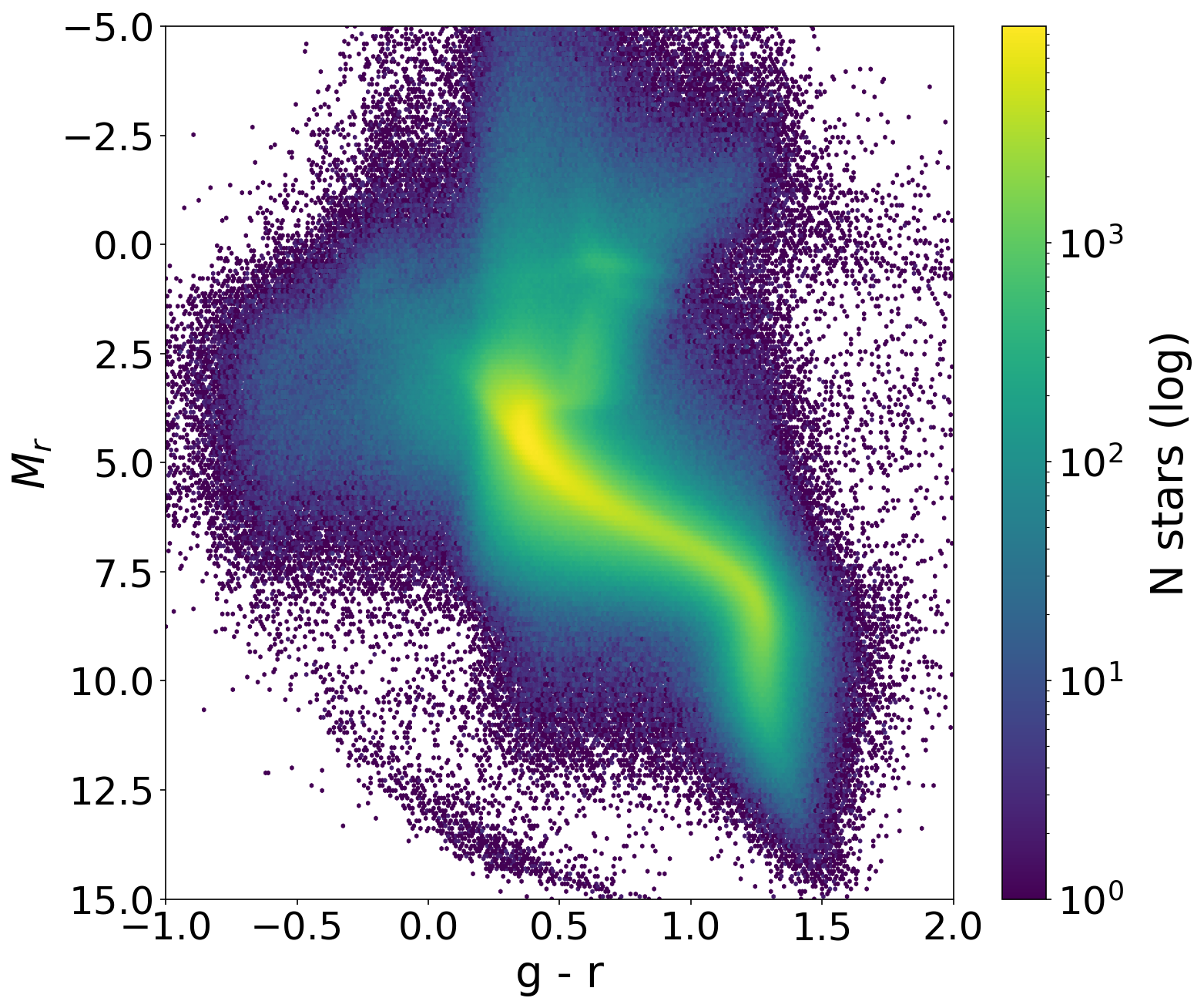}
	\caption{Colour--magnitude diagram of the complete sample for the sources with $\varpi > 0$ ($r \leq 19$~mag). The y-axis is the absolute magnitude in the $r$ band, $M_r$, from the \gaia\ DR3 parallax, and the x-axis the $g-r$ colour; the colour scale shows the number of sources per cell on a logarithmic scale.}
	\label{fig:CMD}
\end{figure}

Among the 96.9 million entries of the dual-mode catalogue of \jplus\ DR4, we retrieved the $3\arcsec$ aperture-corrected photometry and the photometric flags of all the sources with $r \leq 19$~mag and a valid measurement in the 12 bands. We also retrieved the astrometry, photometry, and classification of their \gaia\ DR3 counterparts \citep{2016A&A...595A...1G,GaiaDR3} from the cross-match provided by the archive (Appendix~\ref{Query}), and we removed 2,118 sources due to mismatches with \gaia. Since the Bayesian classification of BANNJOS \citep{Bannjos} was not yet available for DR4, we relied on \gaia\ DR3 to separate stars from extended and extragalactic sources: we kept only the sources with a probability of being a single star of at least 0.95 according to the Discrete Source Classifier \citep[column \texttt{classprob\_dsc\_combmod\_star};][]{Delchambre2023}, which is independent of the \jplus\ photometry. These 14,628,650 stars constitute our complete sample. This is the sample for which we derive stellar parameters. Figure~\ref{fig:JPLUS_DR4} shows the distribution of stars on the sky in Galactic coordinates, and Fig.~\ref{fig:CMD} their colour--magnitude diagram.

The models were trained on a clean subset of the complete sample. We required the photometry to be free of extraction and mask flags in the 12 bands ($\texttt{FLAGS} = 0$ and $\texttt{MASK\_FLAGS} = 0$), and we discarded one bright star ($r < 14$~mag) with an anomalously blue colour, $G - r < -1$~mag, indicative of saturation in the $r$ band. This leaves 10,034,707 stars, which we corrected for Galactic extinction with the 2D map of \citet{2011ApJ...737..103S}, a recalibration of the infrared-based dust map of \citet{1998ApJ...500..525S}.

From the 12 bands we built the two sets of variables used throughout this paper. The first is the set of 11 colours formed by each band and $r$, defined in the sense bluer minus redder band. The second is the set of 11 PCs of the 66 colours that can be formed from the 12 bands (Appendix~\ref{PCA}). We refer to the models fitted with the first set as colour models and to those fitted with the second as PC models.

As a last cleaning step, we iteratively removed the stars lying beyond $5\sigma$ in any of the 11 PCs until no outliers remained. This clipping removes 3,742,545 stars (37\,\%) and leaves 6,292,162. Most of the removed stars are faint ($r_0 > 17$~mag, where $r_0$ is the extinction-corrected $r$ magnitude) or lie outside the colour range of the training sample ($g-r < -0.1$ or $g-r > 1.3$~mag; Fig.~\ref{fig:clipping}), and only 844 (2\,\%) of the stars in common with LAMOST are affected. Note that the clipping is used only to define the training sample; the predictions are computed for the complete sample.

\subsection{LAMOST DR12}\label{sec:spec}

The stellar parameters used to train and test the models come from the medium-resolution survey \citep[MRS; $R \approx 7500$;][]{2020arXiv200507210L} of LAMOST, whose twelfth data release (DR12 v2.0) provides parameters for 3,224,901 co-added spectra. We used the values of \teff, \logg, \feh, and \am\ derived by the LAMOST convolutional neural network (CNN) pipeline, a label-transfer method trained on APOGEE labels that ties the LAMOST parameters to the APOGEE scale \citep[see][for a CNN of this kind applied to the LAMOST MRS]{Wang2020}. We discarded the spectra flagged as affected by moonlight and those without valid CNN parameters or with a non-positive signal-to-noise ratio (\snr), which left 2,484,931 spectra of 1,236,340 unique stars. Among the stars observed more than once, we removed 70 whose spectra had been matched to \gaia\ sources differing by more than 1~mag in $G$, a sign of an ambiguous identification. Since LAMOST does not provide uncertainties for the CNN parameters, we combined the repeated observations of each star into a single measurement with a mean weighted by the squared \snr, $\hat{X} = \sum_{i} (\snr)_i^{2} X_i / \sum_{i} (\snr)_i^{2}$ for $X \in \{\teff, \logg, \feh, \am\}$, which is equivalent to an inverse-variance weighting if the uncertainties scale as the inverse of the \snr. We combined the \snr\ of the individual observations in quadrature, $(\snr)_{\mathrm{comb}} = [\sum_{i} (\snr)_i^{2}]^{1/2}$, and we kept the 788,840 stars with $(\snr)_{\mathrm{comb}} > 30$.

\subsection{Training, validation, and test samples}\label{LAMOST}

We cross-matched our clean photometric sample from J-PLUS with the LAMOST sample using the \jplus\ coordinates and a maximum separation of 0.6\arcsec. To remove mismatches, we kept only the stars with $10 < G < 15$~mag: brighter stars saturate in \jplus\ and were already excluded by the photometric flags, and fainter ones lie beyond the limiting magnitude of the LAMOST MRS \citep{2020arXiv200507210L}, so no genuine counterpart is expected outside this range. This cut removes 44 stars and leaves 42,408 (Fig.~\ref{fig:CMD_L}), which we randomly split into a training sample of 33,926 stars (80\,\%), used to fit the models, and a validation sample of 8,482 stars (20\,\%), used to select them (Sect.~\ref{sec:Polynomial}). We assessed the final performance of the models on an independent test sample of 8,031 stars that were not used at any stage of the modelling.

\section{Method}\label{sec:models}

We modelled each stellar parameter with a polynomial of the \jplus\ colours, or of their PCs, fitted to the training sample, and we estimated the uncertainties of the predictions through a Monte Carlo (MC) scheme. For each parameter we selected the best model for each set of variables, which yields eight models in total. The two families of models yield analogous results. In the rest of the paper we therefore use the colour models, and we present the PC models in Appendix~\ref{PCM}.

\subsection{Polynomial models}\label{sec:Polynomial}

Each stellar parameter $y$ is modelled independently as a polynomial of degree $g$ in $n$ variables $x_1, \dots, x_n$, which are extinction-corrected \jplus\ colours or PCs,
\begin{equation}
\hat{y} = \sum_{k_1 + \dots + k_n \le g} a_{k_1 \dots k_n} \prod_{j=1}^{n} x_j^{k_j},
\label{eq:poly}
\end{equation}
where $\hat{y}$ is the predicted value of $y$ and the sum runs over all the monomials of total degree up to $g$, cross terms included, so that the model has $(n+g)!/(n!\,g!)$ coefficients $a_{k_1 \dots k_n}$. Since the model is linear in the coefficients, we determined them by robust linear regression with the Huber norm, which down-weights the stars with large residuals, as implemented in the \texttt{statsmodels} Python package \citep{seabold2010statsmodels}. The fit also provides the covariance matrix of the coefficients, which we use in Sect.~\ref{sec:uncertainties} to propagate their uncertainties.

To select the variables and the degree of each model, we fitted all the subsets of 2 to 11 variables with polynomials of degree 2 to 5, evaluating every fit on the validation sample with three metrics: the mean squared error (MSE), the coefficient of determination ($R^2$), and the ratio $Q = \mathrm{MSE}/r_{\rm P}^2$, where $r_{\rm P}$ is the Pearson correlation coefficient. The MSE measures the accuracy of the predictions and $r_{\rm P}^2$ how tightly they follow the spectroscopic values; $Q$ decreases when either improves, so we adopted for each parameter and set of variables the model with the lowest $Q$. Before computing the metrics, we removed the validation stars whose residuals exceeded six times the robust standard deviation of the residuals, iterating until convergence or up to five times, so that a few catastrophic outliers do not dominate the comparison between models. This removes between 31 and 45 stars (0.4--0.5\,\%) of the validation sample, depending on the model.

\begin{figure*}[t]
    \centering
    \includegraphics[width=0.9\textwidth]{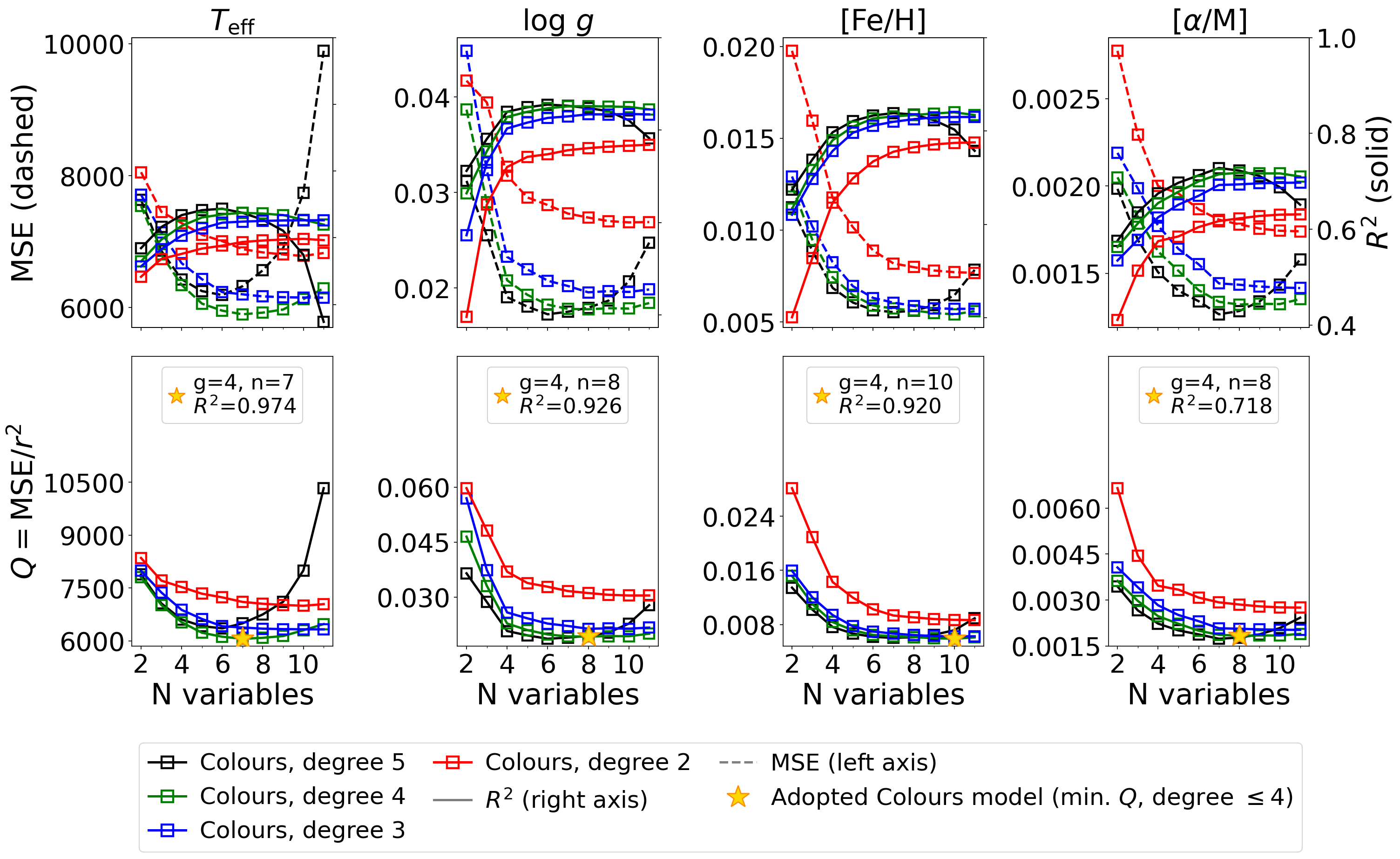}
    \caption{Performance metrics of the best colour model for each number of variables (x-axis) and polynomial degree (degree 2 in red, 3 in blue, 4 in green, and 5 in black). Top row: MSE (dashed lines, left-hand y-axis) and $R^2$ (solid lines, right-hand y-axis). Bottom row: $Q = \mathrm{MSE}/r_{\rm P}^2$. The adopted models, those with the lowest $Q$, are marked with stars; their degree, number of variables, and $R^2$ are given in the legends. }
    \label{fig:metrics_evolution_g}
\end{figure*}

Figure~\ref{fig:metrics_evolution_g} shows, for each degree and each number of variables, the metrics of the best subset of variables. The MSE decreases and $R^2$ increases with the number of variables up to the minimum of $Q$, beyond which additional variables no longer improve the predictions: the models start to overfit the training sample. The degree-5 models degrade faster beyond that point. At their minimum of $Q$ they perform similarly to the degree-4 ones, which we adopted because they are simpler and extrapolate less violently outside the training domain. Table~\ref{tab:best_models} lists the four adopted models, which are marked with stars in Fig.~\ref{fig:metrics_evolution_g}: they use between 7 and 10 variables and have between 330 and 1001 coefficients, that is, roughly 30 to 100 training stars per coefficient.

\begin{table*}[t]
	\centering
	\caption{Adopted polynomial models.}
	\label{tab:best_models}
	\small
	\setlength{\tabcolsep}{4pt}
	\begin{tabular}{lcccccl}
		\hline\hline
		\noalign{\smallskip}
		Parameter & $g$ & $n$ & $N_{\rm coef}$ & $R^2$ & $Q$ & Variables \\
		\noalign{\smallskip}
		\hline
		\noalign{\smallskip}
		\teff & 4 & 7  & 330  & 0.9737 & 6057     & $u - r$, J0378 $-$ $r$, J0395 $-$ $r$, J0430 $-$ $r$, $g - r$, J0515 $-$ $r$, $r$ $-$ J0861 \\
		\logg & 4 & 8  & 495  & 0.9259 & 0.01919  & $u - r$, J0378 $-$ $r$, J0395 $-$ $r$, J0410 $-$ $r$, J0430 $-$ $r$, $g - r$, J0515 $-$ $r$, $r$ $-$ J0861 \\
		\feh  & 4 & 10 & 1001 & 0.9200 & 0.005885 & $u - r$, J0378 $-$ $r$, J0395 $-$ $r$, J0410 $-$ $r$, J0430 $-$ $r$, $g - r$, J0515 $-$ $r$, $r$ $-$ J0660, $r$ $-$ J0861, $r - z$ \\
		\am   & 4 & 8  & 495  & 0.7179 & 0.001837 & $u - r$, J0378 $-$ $r$, J0395 $-$ $r$, J0410 $-$ $r$, J0430 $-$ $r$, J0515 $-$ $r$, $r$ $-$ J0660, $r$ $-$ J0861 \\
		\noalign{\smallskip}
		\hline
	\end{tabular}
		\tablefoot{Polynomial degree ($g$), number of variables ($n$), number of coefficients ($N_{\rm coef}$), $R^2$ and $Q$ on the validation sample, and input colours of each model. The adopted models are marked with stars in Fig.~\ref{fig:metrics_evolution_g}.}
\end{table*}

The variables retained by the selection are informative. Six colours enter the four colour models: those built with $u$, J0378, J0395, J0430, J0515, and J0861, that is, the bands covering the Balmer jump, the Ca\,H\,\&\,K lines, the G band, the Mg\,b triplet, and the Ca triplet. The $g$ band enters all the models except that of \am, the H$\delta$ band (J0410) all except that of \teff, the H$\alpha$ band (J0660) only those of \feh\ and \am, the $z$ band only that of \feh, and the $i$ band none of them. Because the colour models use a small number of explicitly identified colours, they can be applied directly to any survey with the same filter system.

The adopted configuration (2D extinction map, no parallax cut, and photometry alone) was preferred to three alternatives. First, requiring a parallax signal-to-noise ratio of at least 5 (Eq.~\ref{eq:parallax}) for the stars of the clean sample yields a performance within 6\,\% of the adopted one, and the models trained without this cut perform better for the 195 stars that do not pass it (median $Q$ 15\,\% lower). Second, correcting the extinction with the 3D map Bayestar19 \citep{2019ApJ...887...93G}, which requires the parallax cut, performs worse than the 2D map for every parameter on a common validation sample. We attribute this to the distance uncertainties that enter the 3D correction and to the low extinction of the \jplus\ footprint (median $A_r = 0.06$~mag and 90th percentile of 0.17~mag in the training sample). Third, adding the parallax or the distance modulus as a variable yields only a modest improvement in \logg, which does not justify making the predictions dependent on \gaia\ parallaxes.

\subsection{Estimating uncertainties}\label{sec:uncertainties}

We estimated the uncertainties of the predicted parameters of each star with an MC simulation that propagates three sources of error through the models:
\begin{itemize}
\item Photometry. Each magnitude is drawn from a normal distribution centred on its catalogue value, with a dispersion equal to its photometric uncertainty added in quadrature to the calibration precision of its filter (Table~\ref{tab:jplus_dr3_calibration_offsets}). We used the DR3 values of \citet{LopezSanJuan24} as an approximation for DR4, and we treated the calibration term as an additional random error.
\item Extinction. $E(B-V)$ is drawn from a normal distribution with a dispersion of 16\,\% of its value \citep{1998ApJ...500..525S}. Within a realisation, the same $E(B-V)$ is used for all the bands, converted into the extinction of each band with the extinction coefficient of its filter.
\item Model coefficients. Each set of coefficients is drawn from the multivariate normal distribution defined by the covariance matrix of the fit (Sect.~\ref{sec:Polynomial}).
\end{itemize}
We generated 2000 realisations per star. For each of them, we corrected the simulated magnitudes for extinction, computed the colours, and evaluated the models with one realisation of the coefficients, which yields a distribution of 2000 values of each parameter. We adopted the biweight scale \citep{Beers1990} of this distribution, a robust estimate of its dispersion, as the uncertainty of the parameter, and the Spearman correlation coefficient between the distributions of each pair of parameters as their correlation. These correlations arise because a given photometric perturbation shifts several parameters at once. They are not physical relations between the parameters, but they must be taken into account whenever the parameters are combined, and we therefore provide them in the catalogue (Appendix~\ref{sec:correlations}). Note that the MC does not account for the intrinsic scatter of the polynomial relations; in Sect.~\ref{sec:Factor_calibracion} we calibrate the MC uncertainties against the spectroscopic residuals to include this term.

\section{Performance on the test sample}\label{sec:results}

In this section we assess the models on the 8,031 stars of the test sample: first the precision of the predictions, and then the reliability of the MC uncertainties, which we calibrate against the observed residuals.

\subsection{Precision of the predictions}\label{sec:test}

\begin{figure*}[t]
    \centering
    \includegraphics[width=0.9\textwidth]{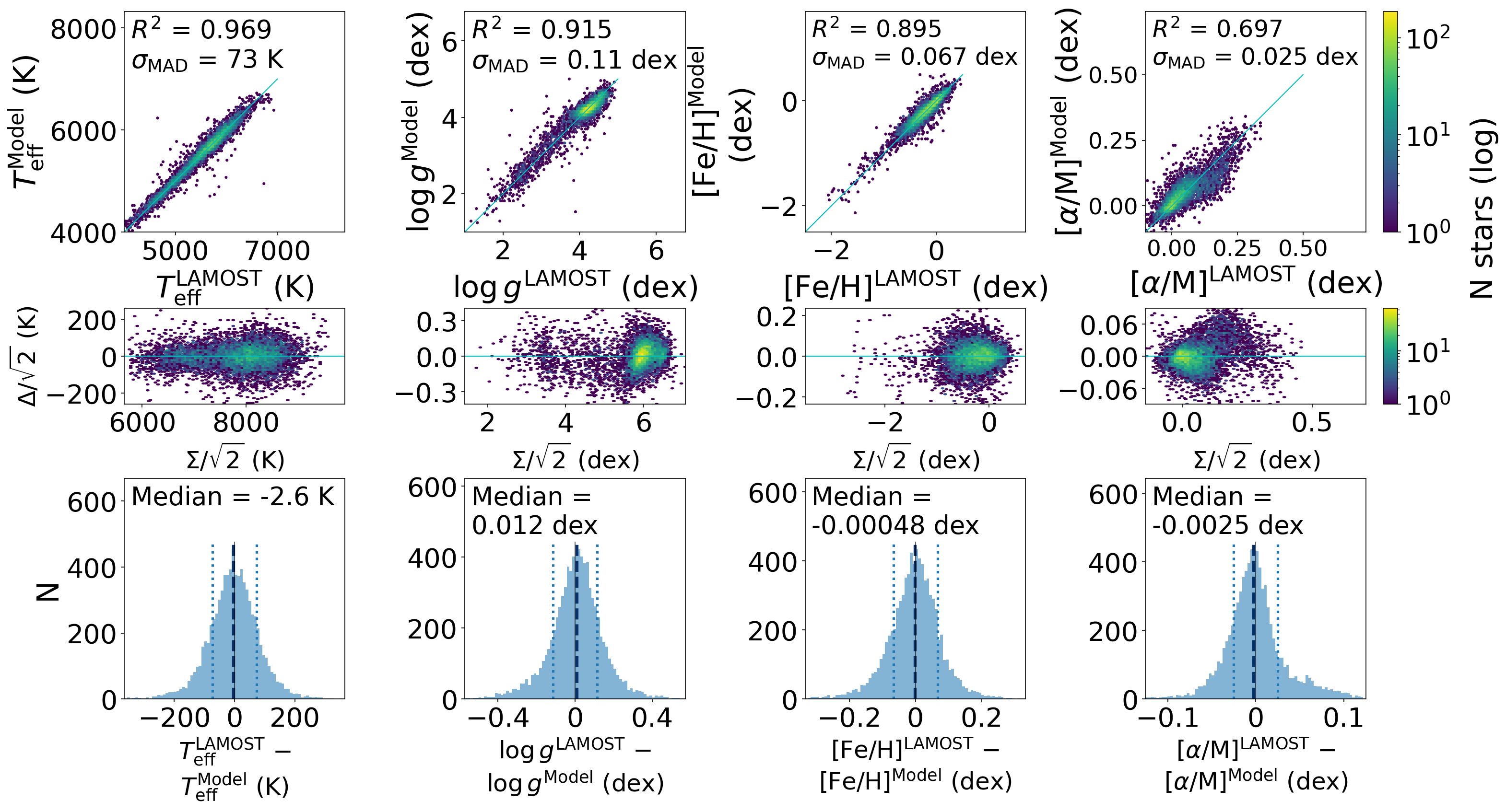}
    \caption{Comparison of the predictions of the colour models with the spectroscopic values from LAMOST DR12 for the test sample (8,031 stars). First row: direct comparison, with the LAMOST values on the x-axis and the predictions on the y-axis; the panels give $R^2$ and the \smad\ of the residuals. Second row: the same comparison rotated by $45^{\circ}$. Third row: distribution of the residuals; the dotted lines mark $\pm\smad$ and the dashed line the median. The colour scale shows the number of stars per cell on a logarithmic scale.}
    \label{fig:BM}
\end{figure*}

Figure~\ref{fig:BM} compares the predicted parameters with the LAMOST values. The first row shows the one-to-one comparison, the second row the same comparison rotated by $45^\circ$, and the third row the distribution of the residuals. The first row of Table~\ref{tab:validation} lists, for each parameter, the median offset and the robust standard deviation of the residuals, $\smad = 1.4826\,\mathrm{MAD}$, where MAD is their median absolute deviation. The offsets are negligible for the four parameters.

\teff\ is the best reproduced parameter, with $R^2 = 0.969$ and $\smad = 73.2$~K, followed by \feh, with $R^2 = 0.895$ and $\smad = 0.067$~dex. This is expected: the broad-band colours are primarily sensitive to the temperature, and the narrow bands J0395 (Ca\,H\,\&\,K) and J0515 (Mg\,b) to the metallicity.

\logg\ is recovered with $R^2 = 0.915$ and $\smad = 0.114$~dex, but the high $R^2$ mainly reflects the separation between dwarfs and giants, which the models place correctly in their respective groups. Within each group, defined by the LAMOST value ($\logg \geq 3.5$ for dwarfs and $\logg < 3.5$ for giants), the correlation is weaker: $R^2 = 0.623$ and $\smad = 0.108$~dex for dwarfs, and $R^2 = 0.707$ and $\smad = 0.225$~dex for giants, which also show a bias of $-0.058$~dex. Giants are poorly represented in the training sample (8.6\,\% of the stars), which explains both their larger scatter and their bias.

\begin{table*}[t]
	\caption{Comparison of the predictions of the colour models with LAMOST and with the external catalogues.}
	\label{tab:validation}
	\centering
	\small
	\setlength{\tabcolsep}{4pt}
	\begin{tabular}{l r rr rr rr rr}
		\hline\hline
		\noalign{\smallskip}
		& & \multicolumn{2}{c}{\teff\ (K)} & \multicolumn{2}{c}{\logg\ (dex)} & \multicolumn{2}{c}{\feh\ (dex)} & \multicolumn{2}{c}{\am\ (dex)} \\
		\cline{3-4}\cline{5-6}\cline{7-8}\cline{9-10}
		\noalign{\smallskip}
		Reference sample & $N$ & $\Delta$ & $\sigma_{\rm MAD}$ & $\Delta$ & $\sigma_{\rm MAD}$ & $\Delta$ & $\sigma_{\rm MAD}$ & $\Delta$ & $\sigma_{\rm MAD}$ \\
		\noalign{\smallskip}
		\hline
		\noalign{\smallskip}
		LAMOST test sample                           & 8,031   & $-2.6$  & 73.2 & $+0.012$ & 0.114 & $0.000$  & 0.067 & $-0.003$ & 0.025 \\
		APOGEE DR17                                  & 17,412  & $+5.0$  & 61.6 & $+0.016$ & 0.156 & $+0.007$ & 0.084 & $+0.016$ & 0.050 \\
		GALAH DR4\tablefootmark{a}                   & 1,637   & $+56.9$ & 71.7 & $+0.056$ & 0.140 & $+0.029$ & 0.089 & $+0.028$ & 0.058 \\
		Huang et al. (2024)\tablefootmark{a}         & 667,673 & $+30.7$ & 61.6 & $+0.059$ & 0.141 & $-0.010$ & 0.082 & $+0.025$ & 0.039 \\
		\noalign{\smallskip}
		\hline
		\noalign{\smallskip}
		APOGEE DR17 $-$ LAMOST DR12\tablefootmark{b} & 38,131  & $-0.6$  & 51.3 & $-0.004$ & 0.091 & $0.000$  & 0.037 & $+0.020$ & 0.032 \\
		GALAH DR4 $-$ LAMOST DR12\tablefootmark{a,b} & 13,786  & $+23.5$ & 80.4 & $+0.002$ & 0.113 & $+0.007$ & 0.058 & $+0.026$ & 0.041 \\
		\noalign{\smallskip}
		\hline
	\end{tabular}
	\tablefoot{Number of stars in common ($N$) and, for each parameter, median offset ($\Delta$, reference minus prediction) and \smad\ of the residuals. The values of the first four rows are those printed in Figs.~\ref{fig:BM}, \ref{fig:Apogee-compare}, \ref{fig:Galha-compare}, and \ref{fig:Huang-compare}.\\
		\tablefoottext{a}{These catalogues provide \afe\ rather than \am; see text.} \tablefoottext{b}{Direct comparison between the two spectroscopic surveys for the stars in common (Fig.~\ref{fig:Apogee-Lamost-Galha}).}}
\end{table*}

\am\ is the most challenging parameter, with $R^2 = 0.697$ and $\smad = 0.025$~dex. The one-to-one comparison in Fig.~\ref{fig:BM} shows the two populations of $\alpha$-poor and $\alpha$-enhanced stars, that is, the thin and thick discs, but the models compress the second one towards the first: the $\alpha$-enhanced stars are predicted with values that are systematically too low, and the gap between the two populations is partly filled. Two effects contribute. The first is regression towards the mean: the training sample is dominated by thin-disc stars, so the fit favours the $\alpha$-poor locus and pulls the minority population towards it. The second is that the $\alpha$-sensitive bands trace individual elements (J0515 the Mg\,b triplet, J0395 and J0861 calcium), whereas \am\ is an average over several elements, so the photometric signal is weaker than for \feh\ and spread over a small dynamic range of 0.4~dex. Part of the scatter is also inherited from the training labels, as the comparison between LAMOST and APOGEE shows (Sect.~\ref{Apogee}).

\subsection{Calibration of the uncertainties}\label{sec:Factor_calibracion}

We tested the MC uncertainties of Sect.~\ref{sec:uncertainties} with the normalised residuals,
\begin{equation}
Z = \frac{y_{\mathrm{LAMOST}} - \hat{y}}{\sqrt{\sigma_{\mathrm{MC}}^2 + \sigma_{\mathrm{LAMOST}}^2}},
\label{eq:zscore}
\end{equation}
where $\sigma_{\mathrm{MC}}$ is the MC uncertainty of the prediction and $\sigma_{\mathrm{LAMOST}}$ the internal precision of the spectroscopic value. Since LAMOST does not provide uncertainties for the CNN parameters, we estimated this precision from the stars with at least two observations with $\snr > 30$, as the median over stars of the \smad\ of their repeated measurements: 34~K, 0.044~dex, 0.018~dex, and 0.0074~dex for \teff, \logg, \feh, and \am, respectively. If the uncertainties are correct, the distribution of $Z$ has a dispersion $\sigma_{Z} \approx 1$; $\sigma_{Z} > 1$ indicates underestimated uncertainties and $\sigma_{Z} < 1$ overestimated ones. We measured $\sigma_{Z}$ by fitting a Gaussian to the histogram of $Z$ (100 bins between the 1st and 99th percentiles).

\begin{figure*}[t]
	\centering
	\includegraphics[width=0.75\textwidth]{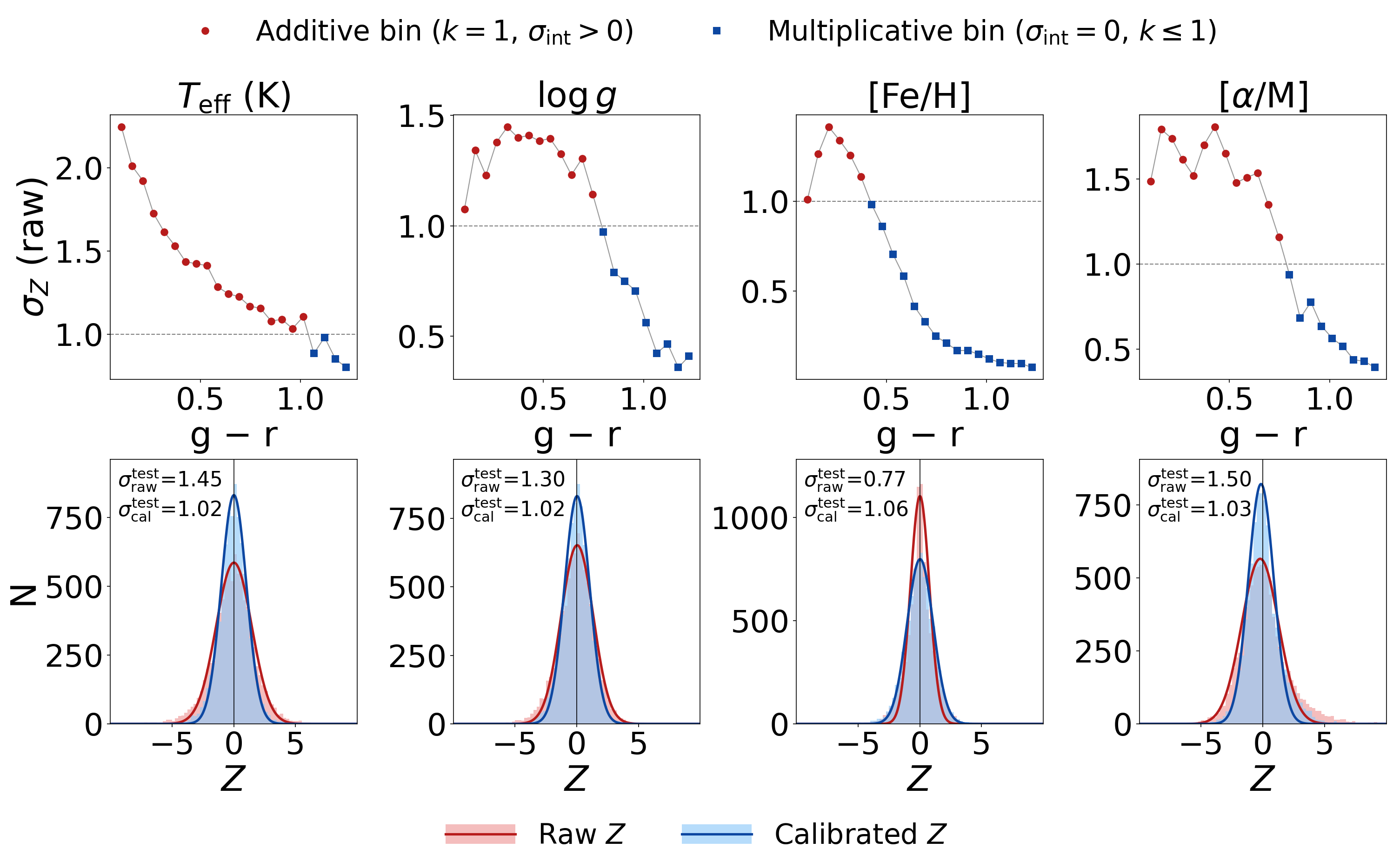}
		\caption{Calibration of the uncertainties of the colour models. First row: $\sigma_{Z}$ of the raw MC uncertainties per $g-r$ bin of the training sample. Red circles mark the bins calibrated with the additive term $\sigma_{\mathrm{int}}$ and blue squares those calibrated with the multiplicative factor $f$ ($k$ in the legend); the line interpolates linearly between bins. Second row: distribution of $Z$ for the test sample with the raw (red) and the calibrated (blue) uncertainties, together with the Gaussian fits from which $\sigma_{Z}$ is measured.}
	\label{fig:Z}
\end{figure*}

For the raw MC uncertainties of the colour models we obtain $\sigma_{Z} = 1.45$, 1.30, 0.77, and 1.50 for \teff, \logg, \feh, and \am, respectively (second row of Fig.~\ref{fig:Z}). The MC uncertainties are therefore underestimated for all the parameters except \feh, as expected from the fact that they do not include the intrinsic scatter of the polynomial relations. Moreover, $\sigma_{Z}$ depends on the $g-r$ colour (first row of Fig.~\ref{fig:Z}): it is largest for the bluest stars (up to 2.2 for \teff) and drops below 1 for the reddest ones.

We corrected for both effects with a hybrid calibration measured on the training sample. We split the training sample into 25 bins of $g-r$, three of which were discarded because they contain very few stars, and we measured $\sigma_{Z}$ in each bin. In the bins with $\sigma_Z > 1$ we added a term $\sigma_{\mathrm{int}}$ in quadrature to the MC uncertainty, and in the bins with $\sigma_Z \leq 1$ we multiplied the MC uncertainty by a factor $f \leq 1$. The calibrated uncertainty of each star is therefore
\begin{equation}
\sigma_{\mathrm{cal}} = \sqrt{\left[f(g-r)\,\sigma_{\mathrm{MC}}\right]^2 + \sigma_{\mathrm{int}}^2(g-r)},
\label{eq:sigcal}
\end{equation}
where $f$ and $\sigma_{\mathrm{int}}$ are interpolated linearly between bins, with $f = 1$ in the additive bins and $\sigma_{\mathrm{int}} = 0$ in the multiplicative ones. For the stars outside the colour range of the training sample, both are kept constant at the value of the nearest edge. By construction, this calibration brings $\sigma_{Z}$ to 1 in every bin of the training sample.

Applied to the test sample, which played no part in the calibration, the calibrated uncertainties give $\sigma_{Z} = 1.02$, 1.02, 1.06, and 1.03 (second row of Fig.~\ref{fig:Z}). Since the calibration depends only on colour, we also checked how $\sigma_Z$ varies with magnitude (Appendix~\ref{app:sigmaZ}): the calibrated $\sigma_Z$ stays between 0.89 and 1.12 over the magnitude range of the test sample, with no common trend with magnitude. Note that the calibrated uncertainties refer to the LAMOST scale and do not include its systematic errors.

\section{The catalogue}\label{sec:catalogue}

We applied the models to the complete sample (Sect.~\ref{sec:jplus}). For each of the 14,628,650 sources, the catalogue provides, for both the colour and the PC models, the four parameters, their calibrated uncertainties (Sects.~\ref{sec:uncertainties} and~\ref{sec:Factor_calibracion}), the six correlation coefficients, and a quality flag per parameter, defined below. The two families of models give equivalent predictions, but the correlations induced by the error propagation are slightly stronger for the PC models (Appendix~\ref{sec:correlations}). We therefore recommend the colour models as the primary product of the catalogue.

\subsection{Validity range and quality flags}\label{sec:flags}

Polynomials extrapolate freely outside the domain in which they were fitted, and the complete sample extends well beyond the training sample in magnitude, colour, and stellar type. We therefore defined for each parameter a validity range from the distribution of the spectroscopic values in the training sample, $[P_{0.1} - 0.2\,\Delta P,\; P_{99.9} + 0.2\,\Delta P]$, where $P_{0.1}$ and $P_{99.9}$ are the 0.1st and 99.9th percentiles and $\Delta P = P_{99.9} - P_{0.1}$ (Table~\ref{tab:validity}). This criterion acts on the output of the models: a star whose colours lie outside the training domain can still receive a value inside the range, so the validity range is a necessary but not a sufficient condition.

\begin{table}[t]
\caption{Validity range of each parameter.}
\label{tab:validity}
\centering
\begin{tabular}{l rr rr}
\hline\hline
\noalign{\smallskip}
Parameter & $P_{0.1}$ & $P_{99.9}$ & Lower & Upper \\
\noalign{\smallskip}
\hline
\noalign{\smallskip}
\teff\ (K)          & 4061     & 6699  & 3533     & 7227  \\
\logg\ (dex)             & 1.523    & 4.827 & 0.862    & 5.488 \\
\feh\ (dex)             & $-1.776$ & 0.336 & $-2.198$ & 0.758 \\
\am\ (dex)       & $-0.099$ & 0.307 & $-0.180$ & 0.388 \\
\noalign{\smallskip}
\hline
\end{tabular}
\tablefoot{Defined from the training sample (33,926 stars) as $[P_{0.1} - 0.2\,\Delta P,\, P_{99.9} + 0.2\,\Delta P]$, with $\Delta P = P_{99.9} - P_{0.1}$.}
\end{table}

The validity range and the uncertainty are complementary criteria, because the MC uncertainty does not capture the extrapolation error: a bright star with small photometric errors that lies outside the training domain receives a small uncertainty together with a meaningless value. We therefore combined the validity range with quality cuts on the calibrated uncertainties, $\sigma_{T_{\mathrm{eff}}} < 350$~K, $\sigma_{\log g} < 0.6$~dex, $\sigma_{\mathrm{[Fe/H]}} < 0.45$~dex, and $\sigma_{[\alpha/\mathrm{M}]} < 0.25$~dex, into one flag per parameter:
\begin{itemize}
\item $\mathrm{flag} = 0$: value inside the validity range and uncertainty below the quality cut;
\item $\mathrm{flag} = 1$: value outside the range and uncertainty above the cut;
\item $\mathrm{flag} = 2$: value inside the range and uncertainty above the cut;
\item $\mathrm{flag} = 3$: value outside the range and uncertainty below the cut, that is, the extrapolated bright stars described above.
\end{itemize}
The quality cuts are permissive: between 56 and 70\,\% of the sources within the validity range of each parameter pass them. Table~\ref{tab:flags} gives the percentage of sources in each flag. We define the recommended sample as the sources with $\mathrm{flag} = 0$ in the four parameters: 4,804,841 sources (32.8\,\%) for the colour models and 4,862,372 (33.2\,\%) for the PC models. The uncertainties of this sample grow steeply beyond $r \approx 15$~mag and reach the quality cuts at $r \approx 17.5$~mag, which is therefore its effective magnitude limit (Appendix~\ref{sec:recommended}). We nevertheless provide the parameters of the complete sample, so that users can define the cuts that suit their science case.

\begin{table}[t]
	\caption{Quality flags of the colour models.}
	\label{tab:flags}
	\centering
	\small
	\setlength{\tabcolsep}{5pt}
	\begin{tabular}{l rrrrr}
		\hline\hline
		\noalign{\smallskip}
		Parameter & In range & flag 0 & flag 1 & flag 2 & flag 3 \\
		\noalign{\smallskip}
		\hline
		\noalign{\smallskip}
		\teff   & 73.14 & 43.50 & 26.85   & 29.65    & 0.00  \\
		\logg   & 62.02 & 35.12 & 37.95    &26.90  & 0.02 \\
		\feh    & 59.91 & 36.81 & 39.81   & 23.11    & 0.28 \\
		\am     & 56.86 & 39.82 & 42.90    & 17.04   &  0.24\\
		\noalign{\smallskip}
		\hline
		\noalign{\smallskip}
		All four & 51.56 & 32.85 & & & \\
		\noalign{\smallskip}
		\hline
	\end{tabular}
	\tablefoot{Percentage of the 14,628,650 \jplus\ DR4 sources within the validity range and in each quality flag. The last row gives the sources that satisfy the condition in all four parameters: 7,543,143 within the validity range and 4,804,841 with flag 0 (the recommended sample).}
\end{table}

\section{Comparison with previous works}\label{sec:comparison}

We compared our predictions with three external catalogues: two spectroscopic surveys, APOGEE DR17 and GALAH DR4, which provide reference values from high-resolution spectra, and the photometric catalogue of \citet{2024ApJ...974..192H}, derived from \jplus\ DR3 with a different method. In the three cases we cross-matched the external catalogue with the recommended sample of the colour models using a maximum separation of 0.6\arcsec, and we excluded the stars of our training sample. Table~\ref{tab:validation} collects the number of stars in common and the median offset and \smad\ of the residuals for each parameter, together with the direct comparisons of LAMOST with APOGEE and with GALAH. The figures are shown in Appendix~\ref{app:comparisons}, which also describes the comparison with GALAH. The two spectroscopic surveys reach only the bright end of the catalogue, whereas the sample of \citet{2024ApJ...974..192H} extends to fainter magnitudes.

\subsection{APOGEE DR17}\label{Apogee}

APOGEE \citep{2017AJ....154...94M, 2022ApJS..259...35A} is a high-resolution ($R \approx 22{,}500$) near-infrared spectroscopic survey that covers all the components of the Milky Way. We kept the stars with $\texttt{SFlag} = 0$, $\texttt{SaFlag} = 0$, and $\snr > 30$ in the VizieR version of the DR17 catalogue (III/286), which leaves 348,420 unique stars, of which 17,412 are in our recommended sample and outside the training sample.

Figure~\ref{fig:Apogee-compare} shows the comparison. The agreement is good for the four parameters, with small offsets (Table~\ref{tab:validation}) and $R^2 = 0.973$, 0.953, 0.890, and 0.572 for \teff, \logg, \feh, and \am, respectively. Two features depart from the one-to-one relation: a zigzag pattern in \logg\ and the deviation of the thick-disc stars in \am. The first is inherited from the training labels: it is absent from the comparison with the LAMOST test sample (Fig.~\ref{fig:BM}) but present in the direct comparison between LAMOST and APOGEE for the stars in common (Fig.~\ref{fig:Apogee-Lamost-Galha}), so our models reproduce the LAMOST \logg\ scale, zigzag included. The second is the compression of the $\alpha$-enhanced population discussed in Sect.~\ref{sec:test}, to which the LAMOST--APOGEE comparison adds an offset of 0.02~dex and a degradation of \am\ at the high-$\alpha$ end that our models also inherit.

To quantify this compression, we selected 2,103 pairs of $\alpha$-rich and $\alpha$-poor stars, identified with Eq.~\ref{eq:alpha_m_feh_cut}, with similar \teff, \logg, and \feh\ (within 100~K, 0.1~dex, and 0.05~dex, respectively), and we compared the difference in \am\ within each pair according to APOGEE and according to our predictions. The predictions recover 23--27\,\% of the difference measured by APOGEE. In other words, the \jplus\ photometry carries information on \am\ beyond its correlation with the other parameters: at fixed \teff, \logg, and \feh, the predictions retain approximately between a quarter and a third of the contrast between the two populations.

\subsection{Photometric parameters from J-PLUS DR3}\label{Huang}

\citet{2024ApJ...974..192H} derived stellar parameters for about five million stars from \jplus\ DR3 with a kernel principal component analysis trained on spectroscopic labels, using colours from \jplus\ and \gaia, whereas our models use \gaia\ only to clean form non-stellar objects our training sample. Like GALAH, they provide \afe\ rather than \am. We selected the stars with quality flags above 0.99 in their catalogue (columns \texttt{FLG\_FEH}, \texttt{FLG\_CFE}, \texttt{FLG\_MGFE}, and \texttt{FLG\_AFE}), which leaves 667,673 stars in common with our recommended sample and outside our training sample. This is by far the largest of the three comparisons, and the only one that reaches the faint end of the recommended sample.

Figure~\ref{fig:Huang-compare} shows the comparison, with $R^2 = 0.968$, 0.896, 0.874, and 0.559 for \teff, \logg, \feh, and \afe, respectively. For \teff\ and \feh\ the two catalogues agree well, with only marginal differences in the tails. These are mostly visible in \teff, for which \citet{2024ApJ...974..192H} give slightly lower values than ours at the low end and slightly higher values at the high end. For \logg, the zigzag pattern appears again, together with a deviation for dwarf stars and a concentration of stars at $\logg \approx 2.5$ in their catalogue that has no counterpart in ours. \afe\ shows the largest scatter relative to its dynamic range, and therefore the lowest $R^2$. This is partly because the comparison reaches fainter stars than the spectroscopic ones, for which the photometric errors of both catalogues are larger. Moreover, since both sets of values are photometric estimates, the comparison cannot tell which of the two catalogues dominates the scatter.

The scatter of the differences increases towards fainter stars, except for \teff: between $r_0 < 15$~mag and $16 \leq r_0 < 17$~mag, \smad\ increases from 0.138 to 0.171~dex in \logg, from 0.074 to 0.126~dex in \feh, and from 0.037 to 0.054~dex in \am. The median offsets in these magnitude bins lie between $+18$ and $+35$~K in \teff, $+0.05$ and $+0.09$~dex in \logg, $-0.032$ and $-0.004$~dex in \feh, and $+0.023$ and $+0.030$~dex in \am. They may be explained by the different label scales and \jplus\ data releases (DR3 and DR4) of the two catalogues and, in the case of \am, by the fact that \citet{2024ApJ...974..192H} provide \afe.

We also used this comparison to examine the uncertainties of the faint stars. We computed $Z$ (Eq.~\ref{eq:zscore}) with the values of \citet{2024ApJ...974..192H} as reference and only our calibrated uncertainties in the denominator, which makes $\sigma_Z$ an upper limit. We find that $\sigma_Z$ decreases towards fainter stars, from 1.1--1.3 at $r_0 < 15$~mag to 0.7--1.0 at $16 \leq r_0 < 17$~mag, that is, our uncertainties become larger than the scatter of the differences. However, both catalogues are based on \jplus\ photometry, so part of their errors is shared and cancels in the differences; this test therefore constrains only the part of our error budget that is not common to both catalogues.

\section{Science verification}\label{sec:applications}

In this section we test the catalogue at the population level: we check that the predicted parameters reproduce the known structure of the Milky Way in the parameter planes, we separate the Galactic components chemically and examine their spatial distribution, and we compare the \feh\ profiles of the predictions with those of LAMOST. Unless stated otherwise, we use the recommended sample of the colour models (Sect.~\ref{sec:flags}).

\subsection{Parameter distributions}\label{sec:param_planes}

\begin{figure}[t]
  \centering
  \includegraphics[width=\columnwidth]{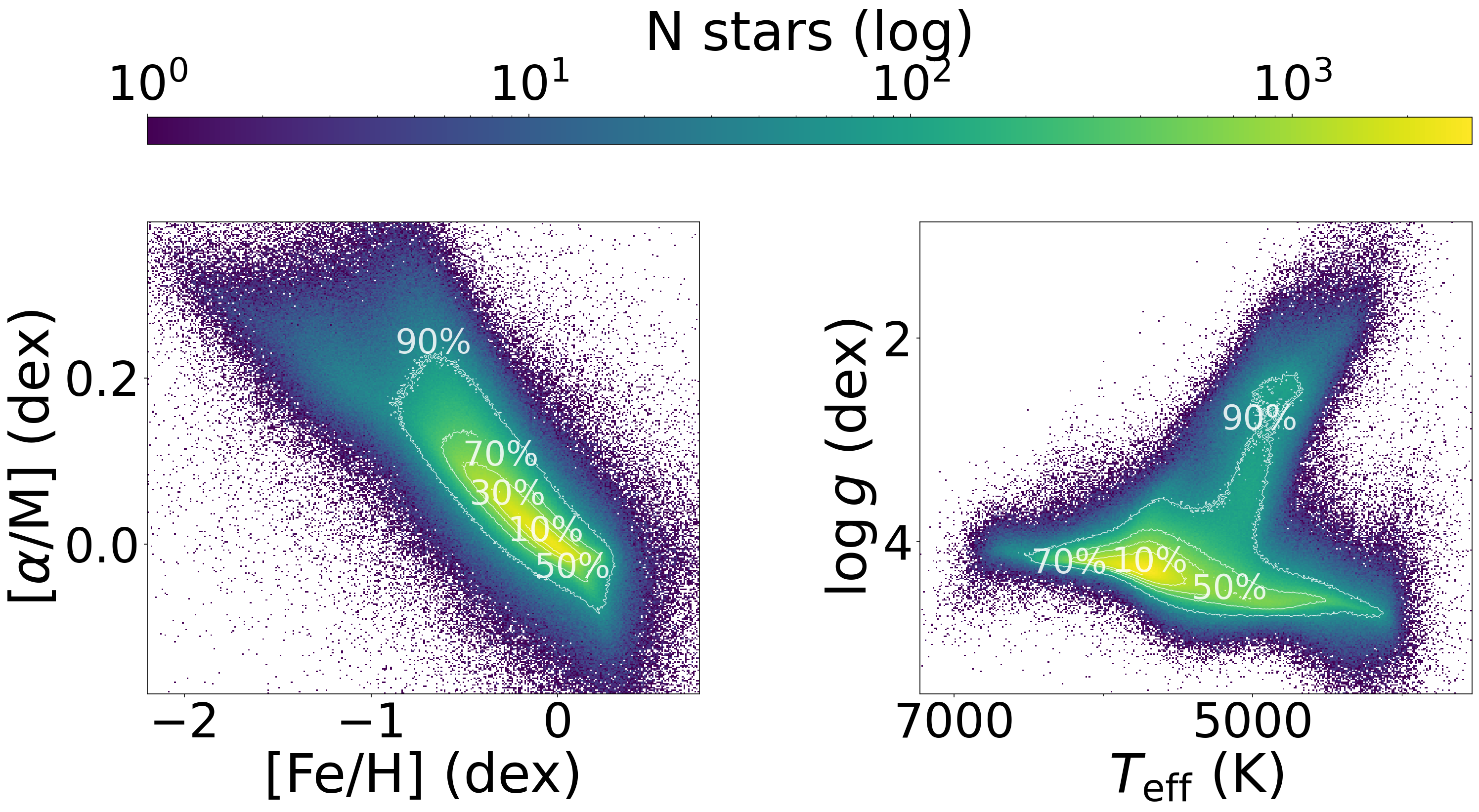}
   \caption{Recommended sample of the colour models in the \feh--\am\ plane (left) and in the \teff--\logg\ plane (right). The colour scale shows the number of stars per cell on a logarithmic scale, and the white contours enclose 30, 50, 70, and 90\,\% of the stars.}
    \label{fig:Predictions}
\end{figure}

Figure~\ref{fig:Predictions} shows the recommended sample in the \feh--\am\ and \teff--\logg\ planes. In the chemical plane, the density contours trace the anticorrelation between \am\ and \feh\ and a change of slope at $\feh \approx -0.7$, where the $\alpha$-enhanced sequence of the thick disc joins the thin-disc sequence. The two sequences are less clearly separated than in the LAMOST sample, as expected from the compression of the $\alpha$-enhanced population (Sect.~\ref{sec:test}) and from the larger uncertainties of the fainter stars. In the Kiel diagram, the main sequence, the red clump at $\teff \approx 4800$~K and $\logg \approx 2.5$, and the giant branch are all recovered.

\subsection{Chemical separation of the Galactic components}\label{Componentes_MW}

We separated the thin disc, the thick disc, and the halo in the \feh--\am\ plane. We classified as halo the stars with $\feh < -1.5$, as thin disc the remaining stars below the line
\begin{equation}
	\am <
	\begin{cases}
		0.1 - 0.061\,\feh, & -1.5 < \feh < 0, \\
		0.1, & \feh \geq 0 ,
	\end{cases}
	\label{eq:alpha_m_feh_cut}
\end{equation}
and as thick disc the rest. This boundary is adapted from that of \citet{2019ApJ...874..102W}, with its slope fitted to the density minimum between the two disc populations. It runs between the modes of the thin and thick discs in the LAMOST MRS DR12 sample with $\snr > 30$, which we located as the centre of the peak cell of a Gaussian-smoothed 2D histogram: $(\feh, \am) = (-0.04, 0.00)$ for the thin disc and $(-0.40, 0.16)$ for the thick disc; the mode of the halo is at $(-1.66, 0.27)$. To limit the cross-contamination between the discs, we further required $\sigma_{[\alpha/\mathrm{M}]} < 0.16$~dex. This leaves 4,287,272 stars down to $r \approx 17.2$~mag, of which 3,786,724, 493,410, and 7,138 are assigned to the thin disc, the thick disc, and the halo, respectively.

To quantify how well the predicted parameters reproduce this classification, we applied the same boundaries to the test sample twice, with the LAMOST parameters and with the predicted ones (Table~\ref{tab:confusion}). Of the stars classified as thin disc from the predictions, 92.4\,\% belong to the thin disc according to LAMOST, and 98.5\,\% of the LAMOST thin-disc stars are recovered. The corresponding purity and completeness are 78.9\,\% and 40.2\,\% for the thick disc, and 91.3\,\% and 95.5\,\% for the halo. The low completeness of the thick disc is a consequence of the compression of the $\alpha$-enhanced population: the thick-disc stars predicted with too low an \am\ fall below the boundary of Eq.~\ref{eq:alpha_m_feh_cut} and are assigned to the thin disc. Since the test sample is brighter than the recommended sample, these values are upper limits for the faint end.

\begin{table}[t]
	\caption{Classification of the stars of the test sample into Galactic components.}
	\label{tab:confusion}
	\centering
	\small
	\setlength{\tabcolsep}{4pt}
	\begin{tabular}{l rrr r}
		\hline\hline
		\noalign{\smallskip}
		& \multicolumn{3}{c}{LAMOST} & \\
		\cline{2-4}
		\noalign{\smallskip}
		Predicted & Thin disc & Thick disc & Halo & Purity (\%) \\
		\noalign{\smallskip}
		\hline
		\noalign{\smallskip}
		Thin disc        & 6,948 & 570 & 1  & 92.4 \\
		Thick disc       & 103   & 384 & 0  & 78.9 \\
		Halo             & 1     & 1   & 21 & 91.3 \\
		\noalign{\smallskip}
		\hline
		\noalign{\smallskip}
		Completeness (\%) & 98.5 & 40.2 & 95.5 & \\
		\noalign{\smallskip}
		\hline
	\end{tabular}
	\tablefoot{Classification with the predicted parameters (rows) versus the LAMOST parameters (columns), using the boundaries of Eq.~\ref{eq:alpha_m_feh_cut} and $\feh = -1.5$, for the stars with $\sigma_{[\alpha/\mathrm{M}]} < 0.16$~dex (8,029 stars).}
\end{table}

To examine the spatial distribution of the components, we computed Galactocentric coordinates with \texttt{astropy} \citep{astropy:2013,astropy:2018,astropy:2022} from the \gaia\ DR3 parallaxes, $\varpi$, for the stars with a parallax signal-to-noise ratio
\begin{equation}
\frac{\varpi - \Delta\varpi}{\sqrt{\sigma_{\varpi}^2 + \sigma_{\mathrm{sys}}^2}} \geq 5,
\label{eq:parallax}
\end{equation}
where $\sigma_{\varpi}$ is the parallax uncertainty, $\Delta\varpi = -0.017$~mas is the median parallax of the quasars in \gaia\ DR3 \citep{2021A&A...649A...4L}, which we adopted as a constant zero point for simplicity, and $\sigma_{\mathrm{sys}} = 0.015$~mas is the systematic uncertainty adopted by \citet{refId0}. We assumed $R_\odot = 8.122$~kpc \citep{2018A&A...615L..15G} and $z_\odot = 20.8$~pc \citep{2019MNRAS.482.1417B} for the position of the Sun.

\begin{figure}[t]
	\centering
	\includegraphics[width=\columnwidth]{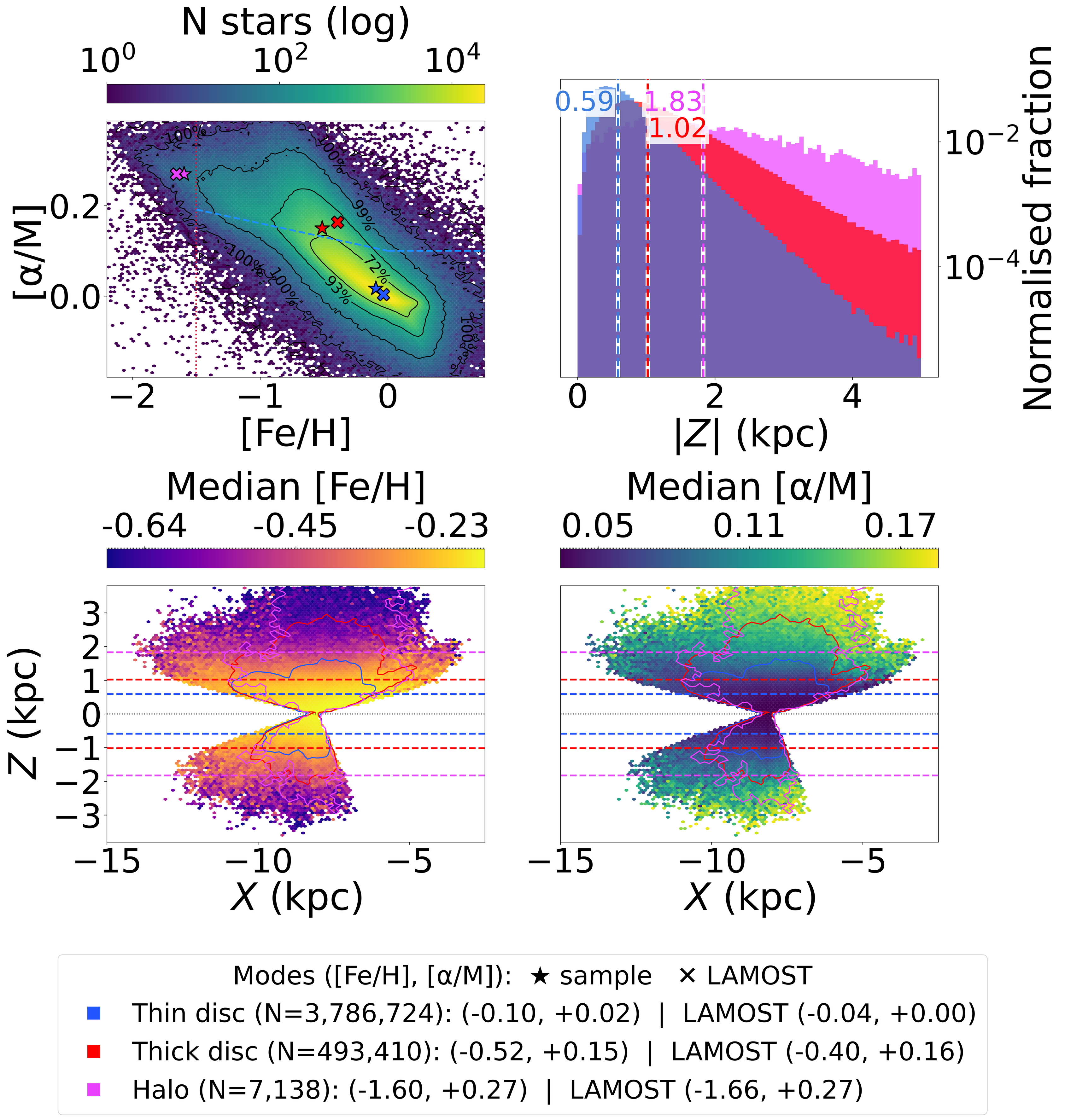}
	\caption{Galactic components in the recommended sample of the colour models with $\sigma_{[\alpha/\mathrm{M}]} < 0.16$~dex (4,287,272 stars). Top left: classification in the \feh--\am\ plane, with the number of stars assigned to each component, density contours labelled with the percentage of stars they enclose, and the modes of each component in the predicted sample (stars) and in LAMOST (crosses). Top right: $|Z|$ distribution of each component, with dashed lines at the medians (in kpc). Bottom: median \feh\ (left) and \am\ (right) in the Galactocentric $X$--$Z$ plane, with the 90\,\% density contour of each component overlaid. The colour scale is histogram-equalised, so that each colour spans the same number of cells.}
	\label{fig:Components}
\end{figure}

Figure~\ref{fig:Components} shows the result for these 4,287,272 stars in all panels. In the chemical plane (top-left panel), the modes of the three components in the predicted sample agree with those of LAMOST to within 0.06~dex in \feh\ and 0.02~dex in \am\ for the thin disc and the halo, whereas the thick-disc mode is 0.12~dex more metal-poor ($\feh = -0.52$ versus $-0.40$), likely because the compression of the $\alpha$-enhanced population moves the metal-rich end of the thick-disc sequence, which lies closest to the boundary, into the thin disc. The three components are ordered in height above the Galactic plane, $|Z|$, as expected, with median values of 0.59, 1.02, and 1.83~kpc for the thin disc, the thick disc, and the halo, respectively (top-right panel). In the Galactocentric $X$--$Z$ plane (bottom panels), solar-metallicity, low-$\alpha$ stars concentrate close to the plane, and the median metallicity decreases and the median \am\ increases with $|Z|$. These trends show that the catalogue recovers the chemo-spatial structure of the Galaxy down to $r \approx 17.5$~mag, about two magnitudes fainter than the stars on which the models were trained ($G < 15$~mag).

\subsection{Metallicity profiles}\label{sec:groups}

\begin{figure}[t]
    \centering
    \includegraphics[width=\columnwidth]{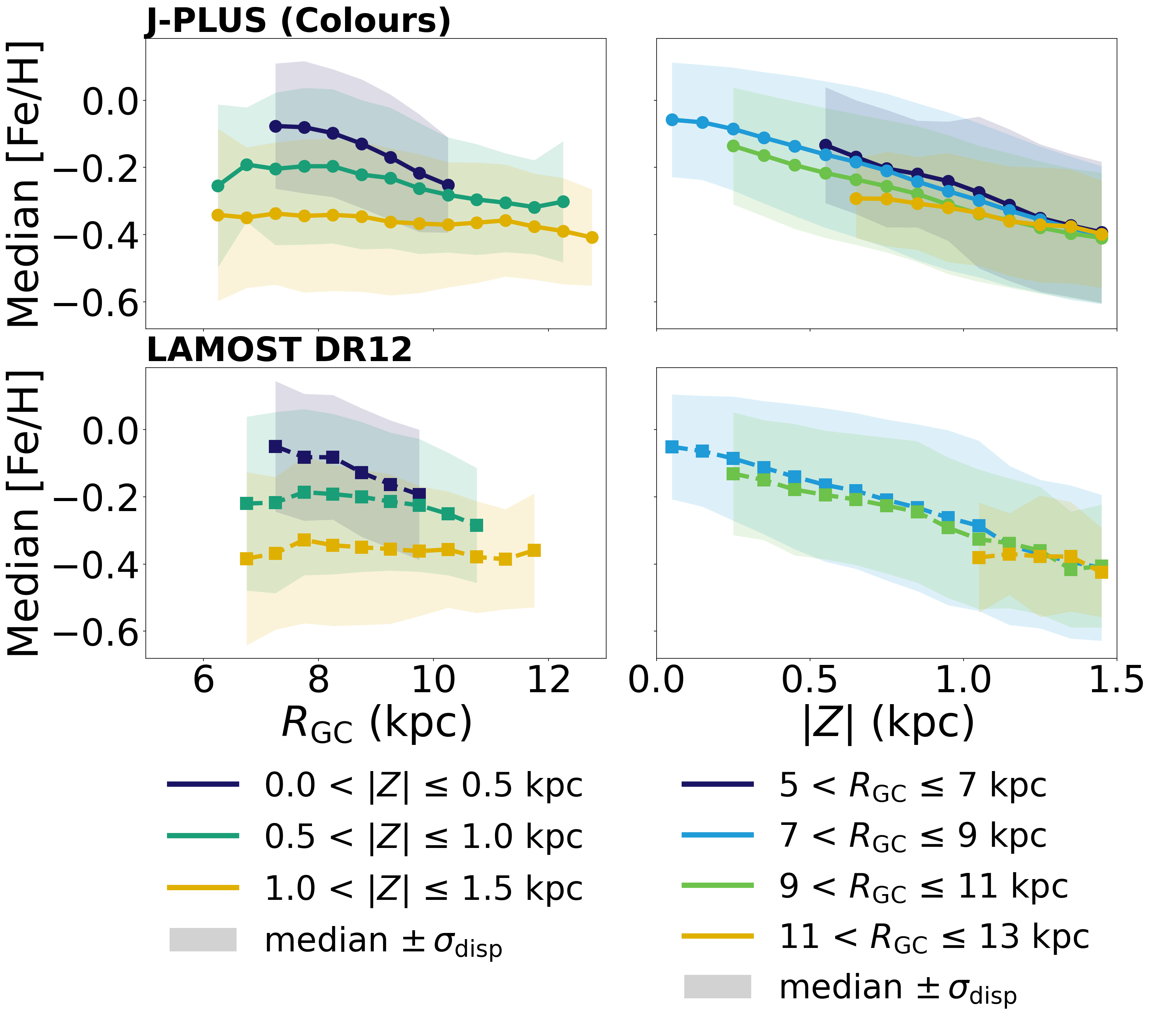}
    \caption{Median \feh\ per bin of Galactocentric radius (left) and height above the Galactic plane (right) for the recommended sample with a parallax signal-to-noise ratio of at least 5 (top panels, solid lines) and for LAMOST DR12 in the same footprint (bottom panels, dashed lines). The shaded regions show the dispersion in each bin, $\sigma_{\rm disp}$ (see text). Only the bins with at least 50 stars are shown.}
    \label{fig:groups}
\end{figure}

We examined the predicted \feh\ as a function of Galactocentric radius, $R_{\rm GC}$, and height, and we compared it with LAMOST DR12. Figure~\ref{fig:groups} shows the median \feh\ per bin of radius and height for the 3,315,308 stars of the recommended sample that satisfy Eq.~\ref{eq:parallax}, and for the 175,838 stars in common between the complete sample and LAMOST DR12, both in the same footprint (the same \jplus\ tiles). The shaded regions show the dispersion in each bin, $\sigma_{\rm disp} = (\sigma_{\rm MAD}^2 - \tilde{\sigma}^2)^{1/2}$, where $\sigma_{\rm MAD}$ is the dispersion of the distribution in the bin and $\tilde{\sigma}$ is the median calibrated uncertainty of our predictions or, for LAMOST, the internal precision of the spectroscopic values (0.018~dex; Sect.~\ref{sec:Factor_calibracion}). For our predictions, $\sigma_{\rm disp}$ is a lower limit to the intrinsic dispersion. 

In the radial profiles, the metallicity decreases with radius in the two bins of height closest to the plane, $0 < |Z| \leq 0.5$~kpc (1,142,940 stars in our sample and 113,097 in LAMOST) and $0.5 < |Z| \leq 1.0$~kpc (1,401,958 and 46,253 stars), as expected from the inside-out formation of the disc \citep{2016ApJ...823...30B}, whereas at $1.0 < |Z| \leq 1.5$~kpc (500,644 and 9,671 stars) it remains approximately constant up to 11~kpc. The dispersion is approximately constant along the profiles, except at the smallest radii, where it drops at $0 < |Z| \leq 0.5$~kpc, probably because of the larger calibrated uncertainties of the fainter stars, and reaches its maximum at $0.5 < |Z| \leq 1.0$~kpc, which indicates a larger diversity of stars in these regions. The LAMOST profiles show the same behaviour over the radii in common.

In the vertical profiles, the metallicity decreases with height in the four bins of radius (31,576, 2,288,490, 704,830, and 20,669 stars), and the four profiles converge at $|Z| \approx 1.4$~kpc, where the thick disc dominates. LAMOST covers only the three bins between 7 and 13~kpc (154,589, 13,952, and 322 stars), in which its profiles behave as ours. Both sets of profiles are consistent with previous studies \citep[e.g.][]{2015RAA....15.1240H,2015RAA....15.1209X,2023ApJ...954..124I}, which found negative radial and vertical metallicity gradients.

\section{Conclusions}\label{sec:conclusions}

We have derived \teff, \logg, \feh, and \am, with their uncertainties and correlations, for the 14,628,650 stars of \jplus\ DR4 with $r \leq 19$~mag. The method is deliberately simple: for each parameter, we fitted a degree-4 polynomial in a subset of the \jplus\ colours, or of their principal components, to 33,926 stars with parameters from the LAMOST MRS. The predictions require only the \jplus\ photometry and an extinction correction, and we make the coefficients and covariance matrices of the models publicly available, so that the parameters can be recomputed and their uncertainties propagated. Our main results are the following:

\begin{itemize}
\item On an independent test sample of 8,031 LAMOST stars, the colour models recover \teff, \logg, \feh, and \am\ with negligible offsets and a precision of $\smad = 73.2$~K, 0.114~dex, 0.067~dex, and 0.025~dex, respectively ($R^2 = 0.969$, 0.915, 0.895, and 0.697); the PC models perform equivalently. The models separate dwarfs from giants, with a precision in \logg\ of 0.11~dex for the former and 0.23~dex for the latter, which are only 8.6\,\% of the training sample.
\item The \jplus\ photometry carries information on \am\ beyond its correlation with the other parameters. The models recover the $\alpha$-poor and $\alpha$-enhanced populations and, at fixed \teff, \logg, and \feh, they retain 23--27\,\% of the difference in \am\ between $\alpha$-rich and $\alpha$-poor stars measured by APOGEE, although the $\alpha$-enhanced population is compressed towards the $\alpha$-poor one.
\item The catalogue provides realistic uncertainties and the correlations between the parameters. After a calibration in colour bins that accounts for the intrinsic scatter of the relations, the MC uncertainties reproduce the residuals of the test sample with $\sigma_Z = 1.01$--1.06.
\item A flag per parameter, based on a validity range and on the calibrated uncertainty, allows the user to select reliable estimates. The recommended sample, with $\mathrm{flag} = 0$ in the four parameters, contains 4,804,841 stars (33\,\%) and reaches $r \approx 17.5$~mag.
\item The predictions agree with APOGEE DR17, GALAH DR4, and the photometric catalogue of \citet{2024ApJ...974..192H}, with median offsets below 60~K and 0.06~dex. The two features that depart from the one-to-one relations, a zigzag pattern in \logg\ and part of the compression of the $\alpha$-enhanced stars, are also present in the direct comparison between LAMOST and the high-resolution surveys, that is, they are inherited from the training labels.
\item The recommended sample reproduces the known structure of the Kiel diagram and of the \feh--\am\ plane, and its metallicity profiles in Galactocentric radius and height follow those of LAMOST. A chemical separation of the thin disc, the thick disc, and the halo, with purities of 92.4, 78.9, and 91.3\,\% and completenesses of 98.5, 40.2, and 95.5\,\% in the test sample, recovers the expected ordering in height above the plane and the vertical gradients of \feh\ and \am\ down to $r \approx 17.5$~mag, about two magnitudes fainter than the stars used to train the models.
\end{itemize}

The catalogue should be used within the limits set by its construction. The predictions are validated spectroscopically for $10 < G < 15$~mag and, against the photometric catalogue of \citet{2024ApJ...974..192H}, down to $r \approx 17$~mag; for fainter stars, the uncertainties rely on the extrapolation in magnitude of the calibration. The flags identify the estimates that lie outside the validity range of the models, which affect about half of the sources in at least one parameter, mostly faint stars with noisy photometry or stellar types not represented in the training sample. They assess the predictions and their uncertainties, not the photometry itself, so they should be combined with the \jplus\ photometric flags. Finally, \am\ is best used in a statistical sense: its scatter is 0.025~dex against LAMOST, on the scale of the training labels, and 0.050 and 0.058~dex against APOGEE and GALAH (Table~\ref{tab:validation}), and its precision is the main factor limiting the chemical separation of the Galactic components.

The method can be easily extended. Retraining the models on the deeper and larger spectroscopic samples now being assembled by DESI, WEAVE, and 4MOST would extend the magnitude range over which the predictions are validated and the stellar types covered by the method. In addition, since the models do not need to be retrained, they can be applied directly to the new sources of the future \jplus\ data releases. Combined with the \gaia\ astrometry, the catalogue provides the chemical information needed to study the structure and history of the Milky Way with a homogeneously selected sample. Overall, this work demonstrates that purely algebraic models, combined with a carefully designed set of filters, can deliver reliable atmospheric parameters and chemical abundances for millions of stars.

\begin{acknowledgements}

Based on observations made with the JAST80 telescope and T80Cam camera for the J-PLUS project at the Observatorio Astrof\'{\i}sico de Javalambre (OAJ), in Teruel, owned, managed, and operated by the Centro de Estudios de F\'{\i}sica del Cosmos de Arag\'on (CEFCA). We acknowledge the OAJ Data Processing and Archiving Department (DPAD; \citealt{upad}) for reducing and calibrating the OAJ data used in this work.

Funding for the J-PLUS Project has been provided by the Governments of Spain and Arag\'on through the Fondo de Inversiones de Teruel; the Aragonese Government through the Research Groups E96, E103, E16\_17R, E16\_20R, E16\_23R, and E16\_26R; grants PID2024-162229NB-I00, PID2024-155572NB-C21, and PID2024-155572NB-C22 (MICIU/AEI/10.13039/501100011033 and ERDF/EU); the Spanish Ministry of Science and Innovation (MCIN/AEI/10.13039/501100011033 y FEDER, Una manera de hacer Europa) with grants PID2021-124918NB-C41, PID2021-124918NB-C42, PID2021-124918NA-C43, and PID2021-124918NB-C44; the Spanish Ministry of Science, Innovation and Universities (MCIU/AEI/FEDER, UE) with grants PGC2018-097585-B-C21 and PGC2018-097585-B-C22; the Spanish Ministry of Economy and Competitiveness (MINECO) under AYA2015-66211-C2-1-P, AYA2015-66211-C2-2, AYA2012-30789, and ICTS-2009-14; and European FEDER funding (FCDD10-4E-867, FCDD13-4E-2685). The Brazilian agencies FINEP, FAPESP, and the National Observatory of Brazil have also contributed to this project.

A.H.P. and A.d.P. acknowledge financial support from grant PID2024-155572NB-C22, and A.H.P. also from grants PID2024-155572NB-C21 and PID2021-124918NB-C41 (see the funding of the J-PLUS project above). A.d.P. and A.A.C. acknowledge the Severo Ochoa grant CEX2021-001131-S (MICIU/AEI/10.13039/501100011033). A.d.P. and B.A. acknowledge the Ram\'on y Cajal fellowships RYC2022-038448-I and RYC2022-037011-I, respectively (MICIU/AEI/10.13039/501100011033, co-funded by the European Social Fund Plus, FSE+). A.d.P. also acknowledges the RyC-MAX grant 20245MAX008 (CSIC) and grant CNS2025-166660, and A.A.C. project PID2023-153123NB-I00 (MCIN/AEI/10.13039/501100011033 and FSE+). J.A. and P.C. acknowledge support from the Spanish Virtual Observatory (\url{https://svo.cab.inta-csic.es}) project, funded by MCIU/AEI/10.13039/501100011033 through grant PID2023-146210NB-I00. J.F.A.'s contribution to this work is part of his academic activities as a professor at Universidad Militar Nueva Granada, Bogot\'a, Colombia. E.M.-P. acknowledges FAPESP grants 2026/11996-0 and 2024/00822-5 (Young Investigator Grant awarded to Felipe Almeida-Fernandes). The work of V.M.P. is supported by NOIRLab, which is managed by the Association of Universities for Research in Astronomy (AURA) under a cooperative agreement with the U.S. National Science Foundation.

Guoshoujing Telescope (the Large Sky Area Multi-Object Fiber Spectroscopic Telescope LAMOST) is a National Major Scientific Project built by the Chinese Academy of Sciences. Funding for the project has been provided by the National Development and Reform Commission. LAMOST is operated and managed by the National Astronomical Observatories, Chinese Academy of Sciences.

Funding for the Sloan Digital Sky Survey IV has been provided by the Alfred P. Sloan Foundation, the U.S. Department of Energy Office of Science, and the Participating Institutions. SDSS-IV acknowledges support and resources from the Center for High Performance Computing at the University of Utah. The SDSS website is \url{www.sdss4.org}. SDSS-IV is managed by the Astrophysical Research Consortium for the Participating Institutions of the SDSS Collaboration including the Brazilian Participation Group, the Carnegie Institution for Science, Carnegie Mellon University, Center for Astrophysics | Harvard \& Smithsonian, the Chilean Participation Group, the French Participation Group, Instituto de Astrof\'isica de Canarias, The Johns Hopkins University, Kavli Institute for the Physics and Mathematics of the Universe (IPMU) / University of Tokyo, the Korean Participation Group, Lawrence Berkeley National Laboratory, Leibniz Institut f\"ur Astrophysik Potsdam (AIP), Max-Planck-Institut f\"ur Astronomie (MPIA Heidelberg), Max-Planck-Institut f\"ur Astrophysik (MPA Garching), Max-Planck-Institut f\"ur Extraterrestrische Physik (MPE), National Astronomical Observatories of China, New Mexico State University, New York University, University of Notre Dame, Observat\'ario Nacional / MCTI, The Ohio State University, Pennsylvania State University, Shanghai Astronomical Observatory, United Kingdom Participation Group, Universidad Nacional Aut\'onoma de M\'exico, University of Arizona, University of Colorado Boulder, University of Oxford, University of Portsmouth, University of Utah, University of Virginia, University of Washington, University of Wisconsin, Vanderbilt University, and Yale University.

This work made use of the Fourth Data Release of the GALAH Survey \citep{2025PASA...42...51B}. The GALAH Survey is based on data acquired through the Australian Astronomical Observatory, under programs: A/2013B/13 (The GALAH pilot survey); A/2014A/25, A/2015A/19, A2017A/18 (The GALAH survey phase 1); A2018A/18 (Open clusters with HERMES); A2019A/1 (Hierarchical star formation in Ori OB1); A2019A/15, A/2020B/23, R/2022B/5, R/2023A/4, R2023B/5 (The GALAH survey phase 2); A/2015B/19, A/2016A/22, A/2016B/10, A/2017B/16, A/2018B/15 (The HERMES-TESS program); A/2015A/3, A/2015B/1, A/2015B/19, A/2016A/22, A/2016B/12, A/2017A/14, A/2020B/14 (The HERMES K2-follow-up program); R/2022B/02 and A/2023A/09 (Combining asteroseismology and spectroscopy in K2); A/2023A/8 (Resolving the chemical fingerprints of Milky Way mergers); and A/2023B/4 (s-process variations in southern globular clusters). We acknowledge the traditional owners of the land on which the AAT stands, the Gamilaraay people, and pay our respects to elders past and present. This paper includes data that has been provided by AAO Data Central (\url{https://datacentral.org.au}).

This work has made use of data from the European Space Agency (ESA) mission \gaia\ (\url{https://www.cosmos.esa.int/gaia}), processed by the \gaia\ Data Processing and Analysis Consortium (DPAC, \url{https://www.cosmos.esa.int/web/gaia/dpac/consortium}). Funding for the DPAC has been provided by national institutions, in particular the institutions participating in the \gaia\ Multilateral Agreement.

This research has made use of \texttt{astropy} \citep{astropy:2013,astropy:2018,astropy:2022}, \texttt{numpy} \citep{2020Natur.585..357H}, \texttt{scipy} \citep{2020NaMet..17..261V}, \texttt{pandas} \citep{mckinney-proc-scipy-2010}, \texttt{statsmodels} \citep{seabold2010statsmodels}, \texttt{TOPCAT} \citep{2005ASPC..347...29T}, and \texttt{matplotlib} \citep{2007CSE.....9...90H}.

\end{acknowledgements}

\bibliographystyle{aa}
\bibliography{JPLUS_DR4_stellar_parameters}

\begin{appendix}
	\renewcommand{\topfraction}{0.95}
	\renewcommand{\bottomfraction}{0.7}
	\renewcommand{\textfraction}{0.05}
	\renewcommand{\floatpagefraction}{0.8}
	\renewcommand{\dbltopfraction}{0.95}
	\renewcommand{\dblfloatpagefraction}{0.8}
	\setcounter{topnumber}{4}
	\setcounter{bottomnumber}{2}
	\setcounter{totalnumber}{6}
	\setcounter{dbltopnumber}{3}

	\section{Archive query}\label{Query}

	Listing~\ref{lst:J-PLUS_query} reproduces the ADQL query used to retrieve the complete sample from the \jplus\ DR4 archive (Sect.~\ref{sec:jplus}): the 3\arcsec\ aperture-corrected magnitudes and their errors, the extinction in each band, the photometric and mask flags, and the \gaia\ DR3 astrometry, photometry, and classification from the cross-match table of the archive.

	\begin{lstlisting}[style=adql, float=t, caption={\jplus\ DR4 archive query used to build the complete sample (abridged: \texttt{X} stands for each of the 12 bands).}, label={lst:J-PLUS_query}]
		SELECT
		MagAB.TILE_ID, MagAB.NUMBER, MagAB.CLASS_STAR,
		MagAB.ALPHA_J2000, MagAB.DELTA_J2000,
		-- for X in uJAVA, J0378, J0395, J0410, J0430, gSDSS,
		--          J0515, rSDSS, J0660, iSDSS, J0861, zSDSS:
		MagAB.MAG_APER_COR_3_0[jplus::X],
		MagAB.MAG_ERR_APER_COR_3_0[jplus::X],
		MagAB.FLAGS[jplus::X], MagAB.MASK_FLAGS[jplus::X],
		ext.ax[jplus::X],
		gaia.RAdeg, gaia.DEdeg, gaia.Plx, gaia.e_Plx,
		gaia.PM, gaia.pmRA, gaia.e_pmRA, gaia.pmDE, gaia.e_pmDE,
		gaia.RV, gaia.e_RV, gaia.Gmag, gaia.PSS
		FROM jplus.MagABDualObj AS MagAB
		INNER JOIN jplus.MWExtinction AS ext
		ON (ext.TILE_ID = MagAB.TILE_ID AND ext.NUMBER = MagAB.NUMBER)
		LEFT JOIN jplus.xmatch_gaia_dr3 AS gaia
		ON (gaia.TILE_ID = MagAB.TILE_ID AND gaia.NUMBER = MagAB.NUMBER)
		WHERE
		MagAB.MAG_APER_COR_3_0[jplus::X] < 99  -- all bands except rSDSS
		AND MagAB.MAG_APER_COR_3_0[jplus::rSDSS] <= 19
	\end{lstlisting}
	\section{Sample selection}\label{app:sample}
	\subsection{The complete sample}\label{app:complete}

\begin{figure}[t]
    \centering
    \includegraphics[width=\columnwidth]{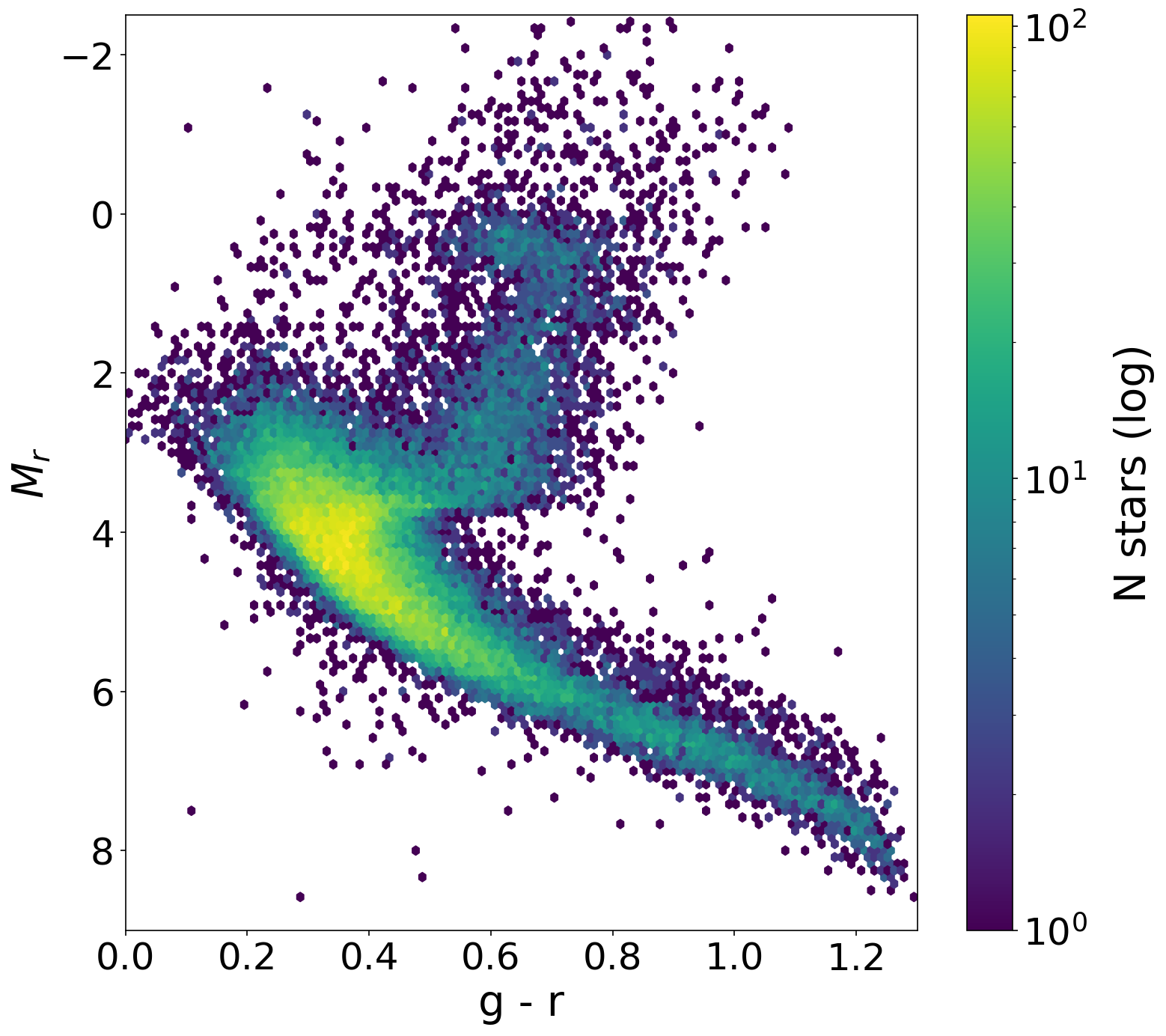}
    \caption{Colour--magnitude diagram of the cross-matched \jplus\ DR4--LAMOST DR12 sample (42,408 stars); only the stars with $\varpi > 0$ are shown. The y-axis is the absolute magnitude in the $r$ band, $M_r$, and the x-axis the $g-r$ colour; the colour scale shows the number of stars per cell on a logarithmic scale.}
    \label{fig:CMD_L}
	\end{figure}

	Figure~\ref{fig:CMD} shows the colour--magnitude diagram of the complete sample for the sources with a positive \gaia\ parallax. The main sequence, the red clump, and the giant branch are clearly visible, together with the white-dwarf sequence at $g - r \approx -0.3$ and $M_r \approx 11$--13~mag and a population of blue sources at $g - r < 0$, which includes hot main-sequence stars and probably blue horizontal-branch stars and misclassified extragalactic sources. Among the stars with a parallax signal-to-noise ratio of at least 5 (Eq.~\ref{eq:parallax}) and no \jplus\ flags in any filter, there are 2,928 white dwarfs, 2,751 hot main-sequence stars, and 2,210 blue horizontal-branch stars. None of these populations is represented in the training sample (Fig.~\ref{fig:CMD_L}), which is one of the reasons behind the quality flags of Sect.~\ref{sec:flags}.

	\subsection{Principal component analysis}\label{PCA}

\begin{figure}[!htbp]
		\centering
		\includegraphics[width=\columnwidth]{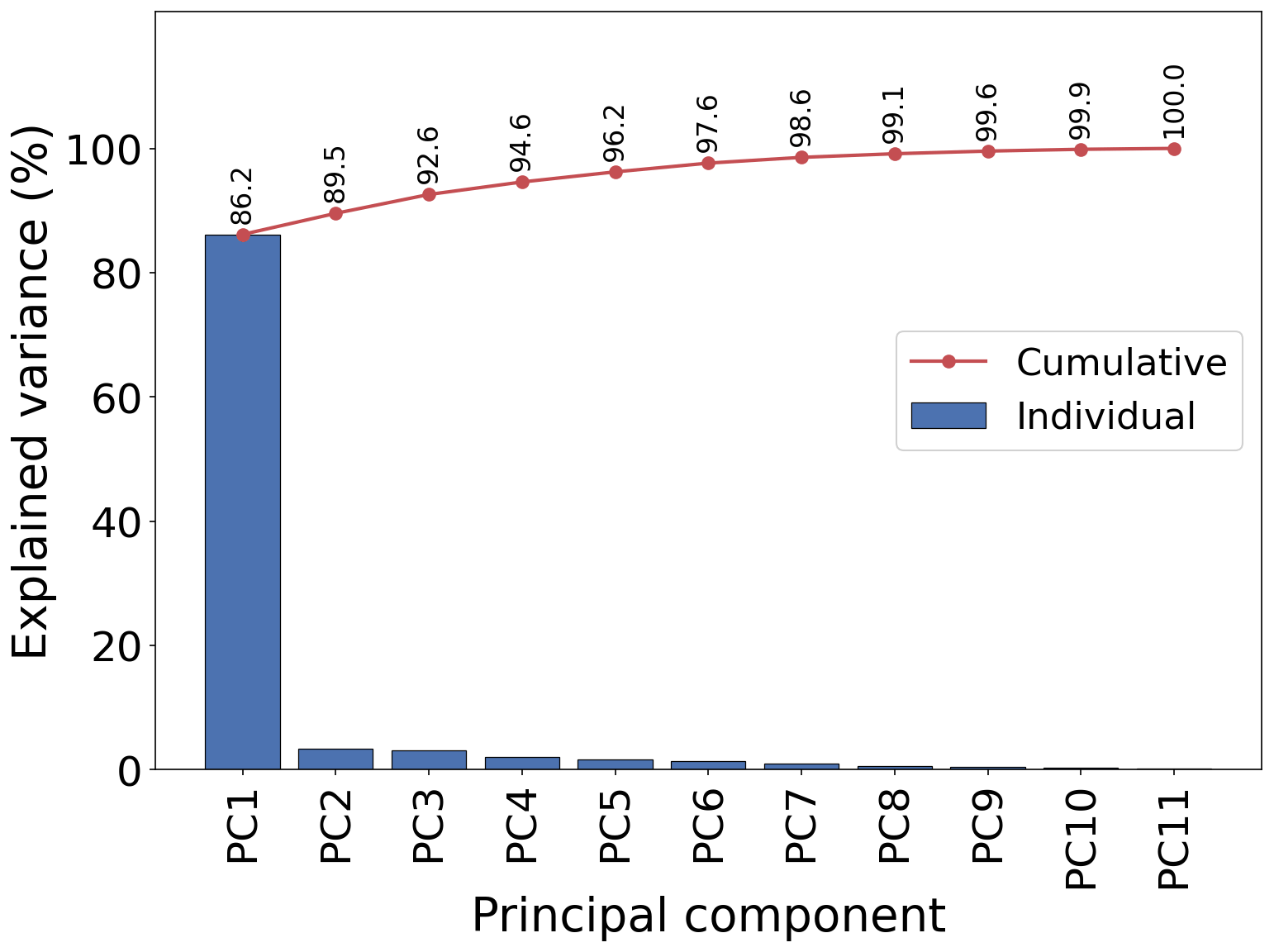}
		\caption{Fraction of the variance of the 66 colours explained by each principal component (bars) and cumulative fraction (line and labels).}
		\label{fig:VE}
	\end{figure}

	We performed the PCA on the 66 extinction-corrected colours of the clean sample. Since the 12 bands provide only 11 linearly independent colours, 11 components reproduce the colours exactly, and the PCA is a rotation of the colour space rather than a compression of it. The variance is nevertheless very unevenly distributed (Fig.~\ref{fig:VE}): PC1 accounts for 86.2\,\% of it, PC1 to PC3 for 92.6\,\%, PC1 to PC7 for 98.6\,\%, and PC8 to PC11 for the remaining 1.4\,\%. The components of lowest variance still carry information on the stellar parameters (Appendix~\ref{PCM}), so we kept all 11 as candidate variables.

\begin{figure}[!htbp]
		\centering
		\includegraphics[width=\columnwidth]{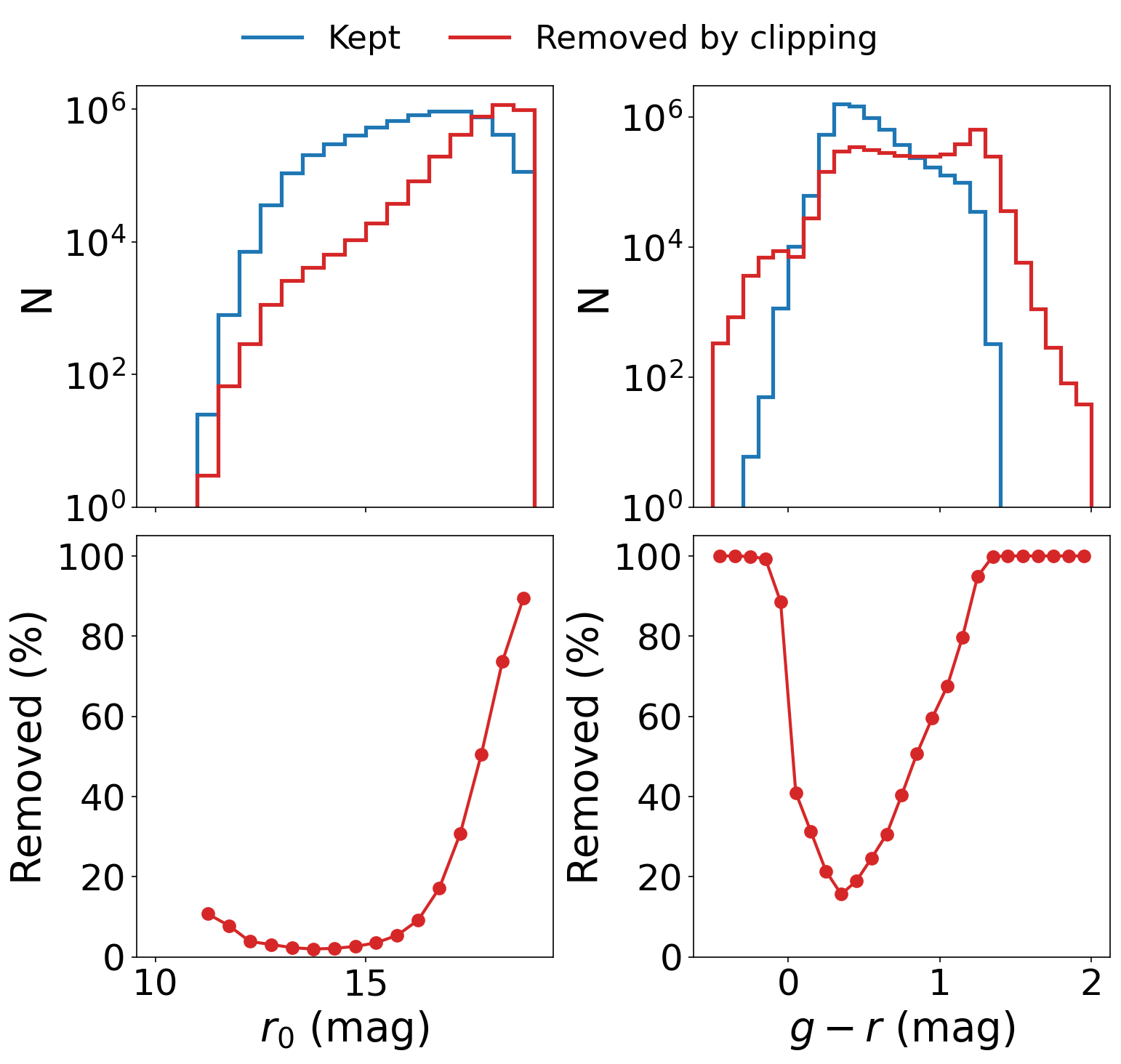}
		\caption{Sources removed by the iterative $5\sigma$ clipping in PC space (red) and kept (blue) in the clean sample, as a function of the extinction-corrected $r$ magnitude (left) and of the $g-r$ colour (right). Top: number of sources per bin; bottom: fraction removed per bin.}
		\label{fig:clipping}
	\end{figure}

	The iterative $5\sigma$ clipping of Sect.~\ref{sec:jplus} was applied in this space, and it converged after 18--23 iterations, depending on the version of the clean sample (Sect.~\ref{sec:Polynomial}). Figure~\ref{fig:clipping} shows the sources that it removes from the adopted clean sample as a function of magnitude and colour. The removed fraction is below 10\,\% between $r_0 \approx 11.5$ and 16.5~mag, but it rises to about 50\,\% at $r_0 \approx 17.8$~mag and to 90\,\% at $r_0 \approx 18.8$~mag. In colour, it is lowest (15\,\%) at $g-r \approx 0.35$~mag, grows steadily towards the red, to 50\,\% at $g-r \approx 0.85$~mag, and reaches 100\,\% for $g-r < -0.1$ and $g-r > 1.3$~mag, that is, outside the colour range of the training sample (Fig.~\ref{fig:CMD_L}). The clipping therefore removes the faint sources, whose colours are dominated by photometric noise, and the sources bluer or redder than any star of the training sample, among them the white dwarfs and the M dwarfs.

	We preferred the iterative clipping to a fixed criterion (a single $5\sigma$ clipping pass) because, with the latter, the information on the stellar parameters is more diluted among the PCs: with the iterative criterion, PC3 carries 0.52~nats of mutual information on \feh\ and 0.29~nats on \teff, whereas with a fixed criterion no PC separates them (0.44 and 0.45~nats, respectively, for PC2). 

	\section{External comparisons}\label{app:comparisons}

\begin{figure*}[t]
    \centering
    \includegraphics[width=0.9\textwidth]{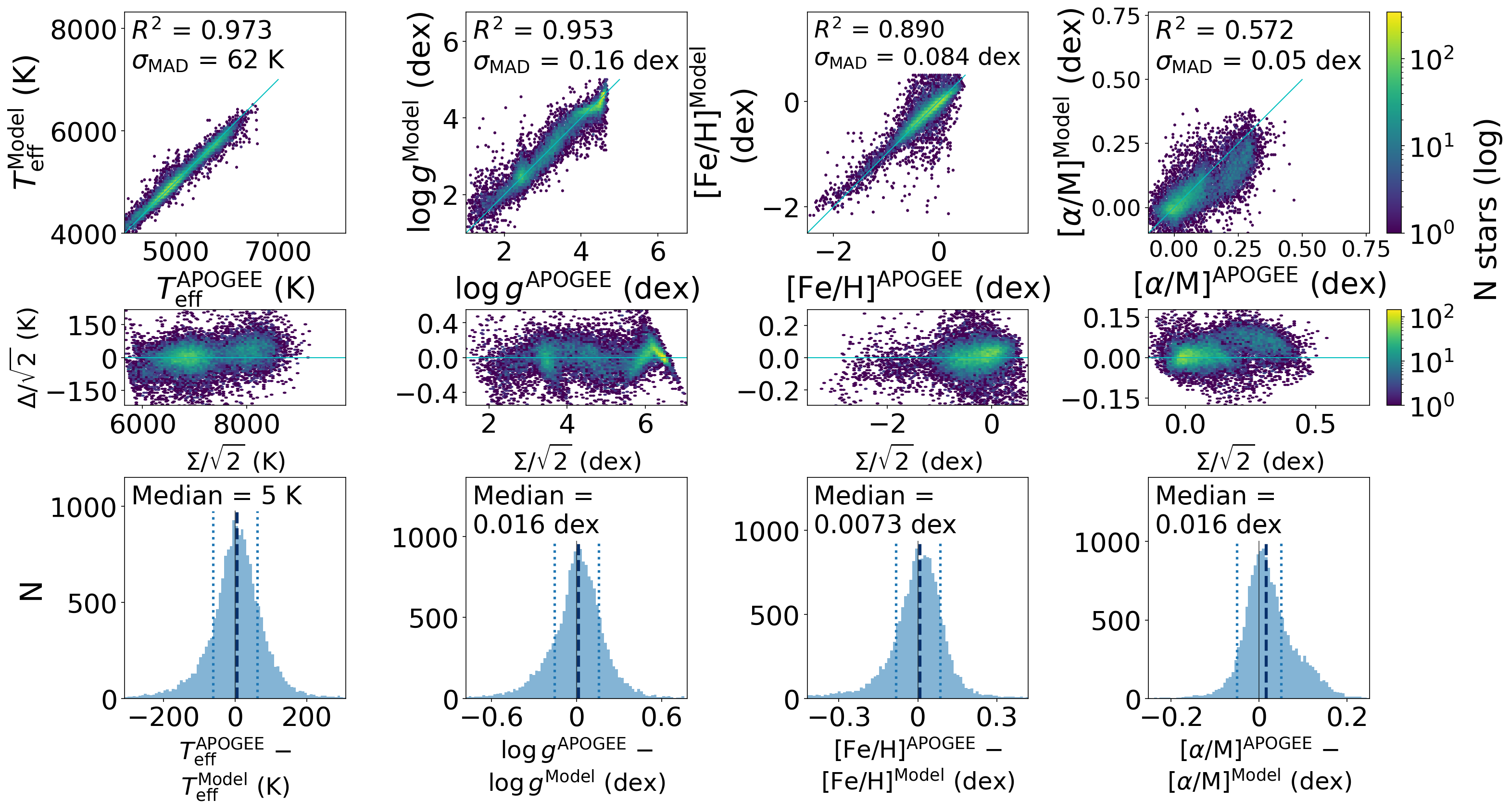}
    \caption{Comparison of the stellar parameters predicted by the colour models with those of APOGEE DR17 (17,412 stars). Panels as in Fig.~\ref{fig:BM}.}
    \label{fig:Apogee-compare}
	\end{figure*}

\begin{figure*}[!tp]
		\centering
		\includegraphics[width=\textwidth]{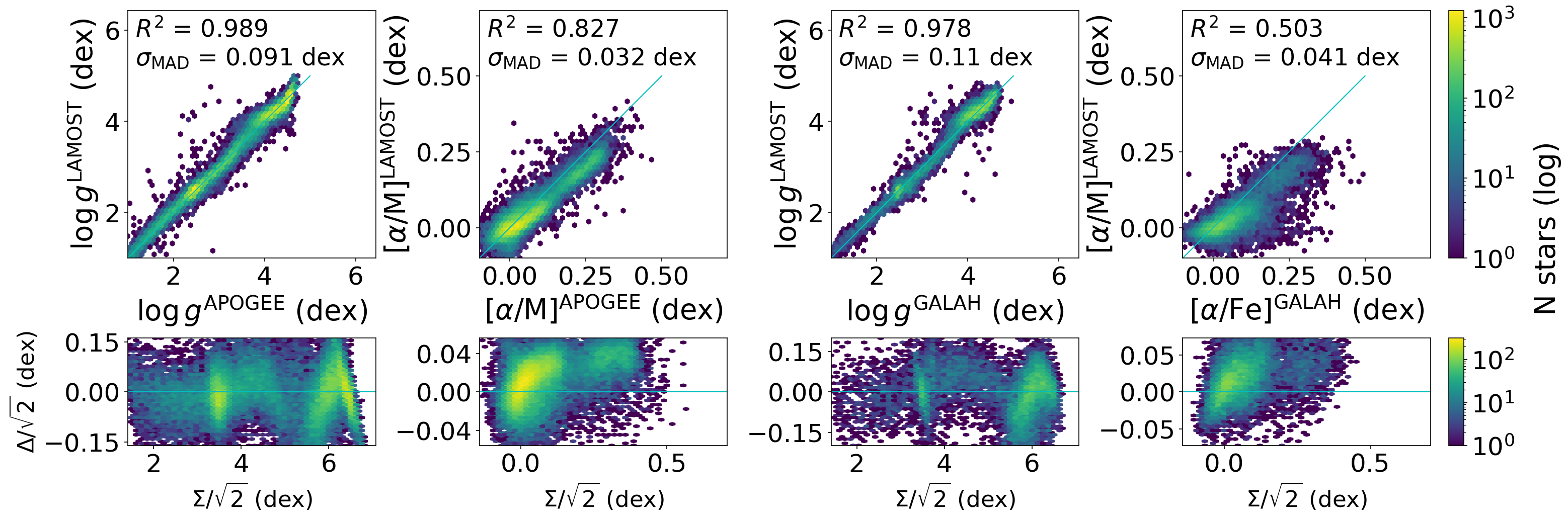}
		\caption{Comparison of the \logg\ and \am\ of LAMOST DR12 with those of APOGEE DR17 (38,131 stars in common; left) and of GALAH DR4 (13,786 stars; right). First row: direct comparison, with $R^2$ and the \smad\ of the residuals. Second row: the same comparison rotated by $45^{\circ}$. The GALAH \afe\ is defined as in Fig.~\ref{fig:Galha-compare}.}
		\label{fig:Apogee-Lamost-Galha}
	\end{figure*}

\begin{figure*}[!tp]
		\centering
		\includegraphics[width=0.8\textwidth]{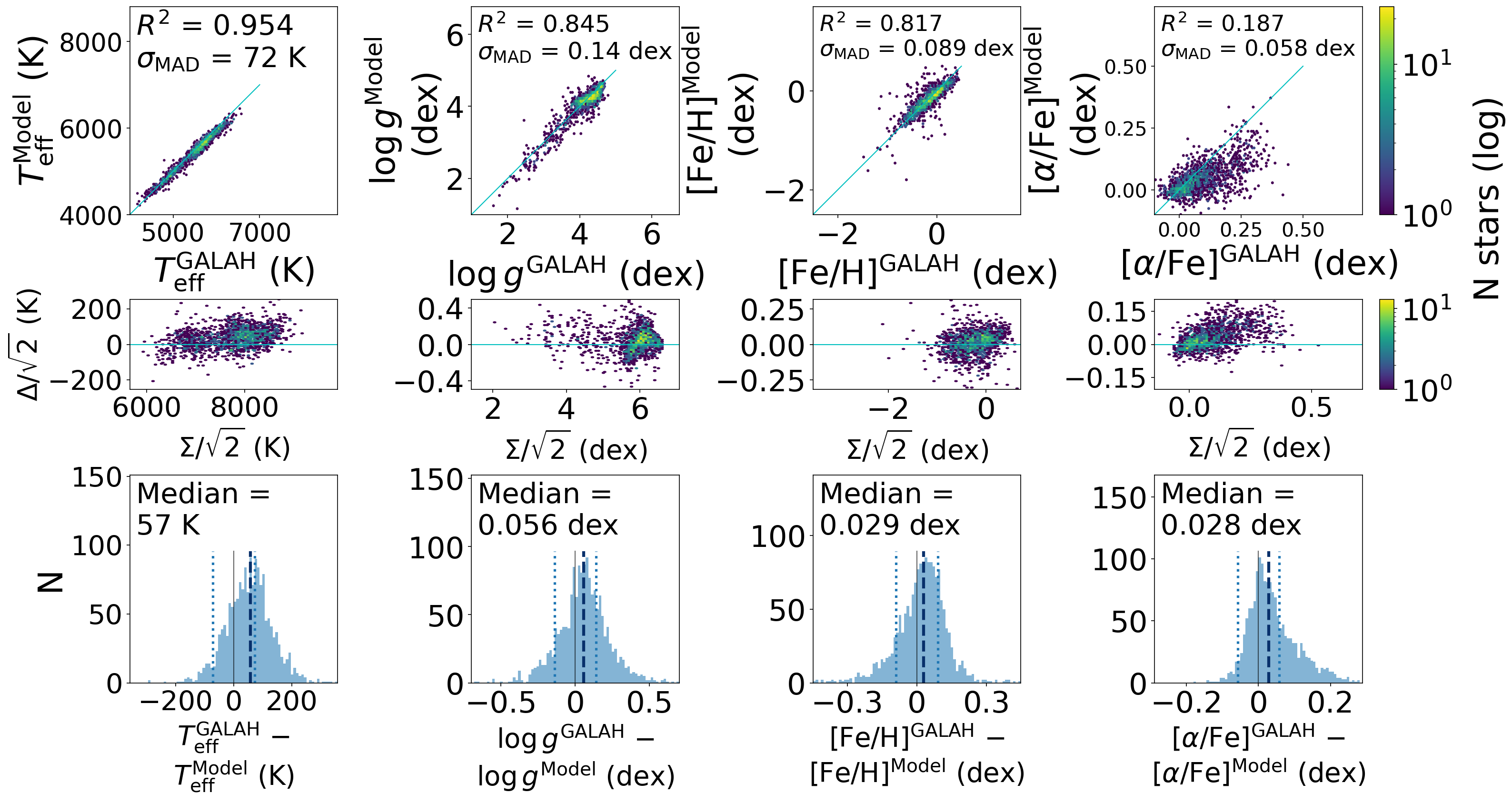}
		\caption{Comparison of the stellar parameters predicted by the colour models with those of GALAH DR4 (1,637 stars). Panels as in Fig.~\ref{fig:BM}; the GALAH \afe\ is the median of [O/Fe], [Mg/Fe], [Si/Fe], [Ca/Fe], and [Ti/Fe].}
		\label{fig:Galha-compare}
	\end{figure*}

\begin{figure*}[!tp]
		\centering
		\includegraphics[width=0.8\textwidth]{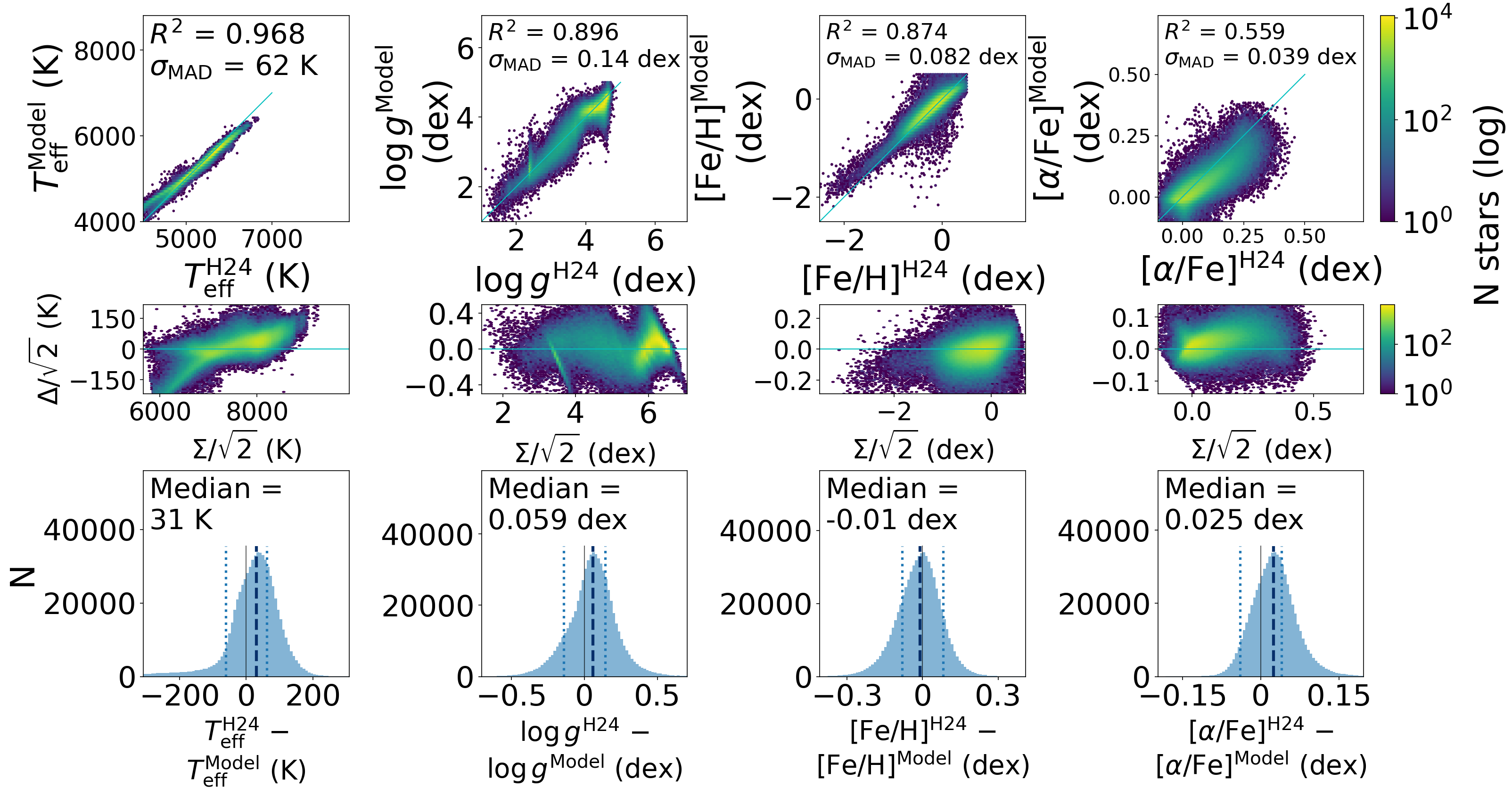}
		\caption{Comparison of the stellar parameters predicted by the colour models with those of \citet{2024ApJ...974..192H} (667,673 stars). Panels as in Fig.~\ref{fig:BM}.}
		\label{fig:Huang-compare}
	\end{figure*}

	This appendix shows the comparisons discussed in Sect.~\ref{sec:comparison}: those of the colour models with APOGEE DR17 (Fig.~\ref{fig:Apogee-compare}), with GALAH DR4 (Fig.~\ref{fig:Galha-compare}), and with the photometric catalogue of \citet{2024ApJ...974..192H} (Fig.~\ref{fig:Huang-compare}), and the direct comparisons of the LAMOST DR12 parameters with those of APOGEE DR17 and GALAH DR4 for the stars in common (Fig.~\ref{fig:Apogee-Lamost-Galha}), which show that the zigzag pattern in \logg\ and part of the \am\ scatter are inherited from the training labels. The statistics printed in the panels are collected in Table~\ref{tab:validation}.

	\subsection{GALAH DR4}\label{Galah}

GALAH \citep{2015MNRAS.449.2604D, 2025PASA...42...51B} is a high-resolution ($R \approx 28{,}000$) optical spectroscopic survey whose fourth data release contains 1,085,520 spectra of 917,588 stars, with stellar parameters and abundances of many individual elements. Since GALAH does not provide \am, we used as an approximation the median of [O/Fe], [Mg/Fe], [Si/Fe], [Ca/Fe], and [Ti/Fe], which we denote \afe; including O increases the $R^2$ of the comparison from 0.136 to 0.187. We kept the stars with $\texttt{flag\_sp} = 0$, no flags in \feh\ or in the five abundances above, and $\snr > 30$, which leaves 501,200 stars, of which 1,637 are in our recommended sample and outside the training sample.

Figure~\ref{fig:Galha-compare} shows the comparison. Most stars lie close to the one-to-one line, with offsets of 57~K in \teff\ and 0.03--0.06~dex in the other parameters (Table~\ref{tab:validation}), and $R^2 = 0.954$, 0.845, 0.817, and 0.187 for \teff, \logg, \feh, and \afe, respectively. The two features of the APOGEE comparison are also present: the zigzag pattern in \logg\ and the deviation of the thick-disc stars in \afe, which here also produces an asymmetric distribution of the residuals, skewed towards $\alpha$-enhanced values. The very low $R^2$ of \afe\ reflects, in addition to the compression discussed in Sect.~\ref{sec:test}, the fact that the two quantities are not the same: our models predict \am\ on the LAMOST scale, whereas the GALAH value is a median of five abundance ratios. Both features are also present in the direct comparison between LAMOST and GALAH for the 13,786 stars in common (Fig.~\ref{fig:Apogee-Lamost-Galha}; last row of Table~\ref{tab:validation}), in which the difference between the two $\alpha$ scales grows towards the $\alpha$-enhanced end (median offset of 0.026~dex). As with APOGEE, this confirms that both features are inherited from the training labels.
	\section{Monte Carlo uncertainties}\label{app:mc}

	\subsection{Calibration precision}\label{app:calib}

	Table~\ref{tab:jplus_dr3_calibration_offsets} lists the calibration precision of each \jplus\ filter, added in quadrature to the photometric errors in the MC simulation of Sect.~\ref{sec:uncertainties}. The values are those of DR3 \citep{LopezSanJuan24}, which we used as an approximation for DR4.

	\begin{table}[!htbp]
		\centering
		\caption{\jplus\ DR3 internal calibration precision for each filter \citep{LopezSanJuan24}.}\label{tab:jplus_dr3_calibration_offsets}
		\begin{tabular}{lc}
			\hline\hline
			Filter & Calibration precision [mmag] \\
			\hline
			$u$   & 10.3 \\
			J0378 & 9.1 \\
			J0395 & 8.7 \\
			J0410 & 4.7 \\
			J0430 & 4.7 \\
			$g$   & 2.8 \\
			J0515 & 3.5 \\
			$r$   & 2.5 \\
			J0660 & 2.9 \\
			$i$   & 2.4 \\
			J0861 & 3.3 \\
			$z$   & 3.2 \\
			\hline
		\end{tabular}
	\end{table}

	\subsection{Calibrated uncertainties as a function of magnitude and colour}\label{app:sigmaZ}

\begin{figure*}[!tp]
		\centering
		\includegraphics[width=0.8\textwidth]{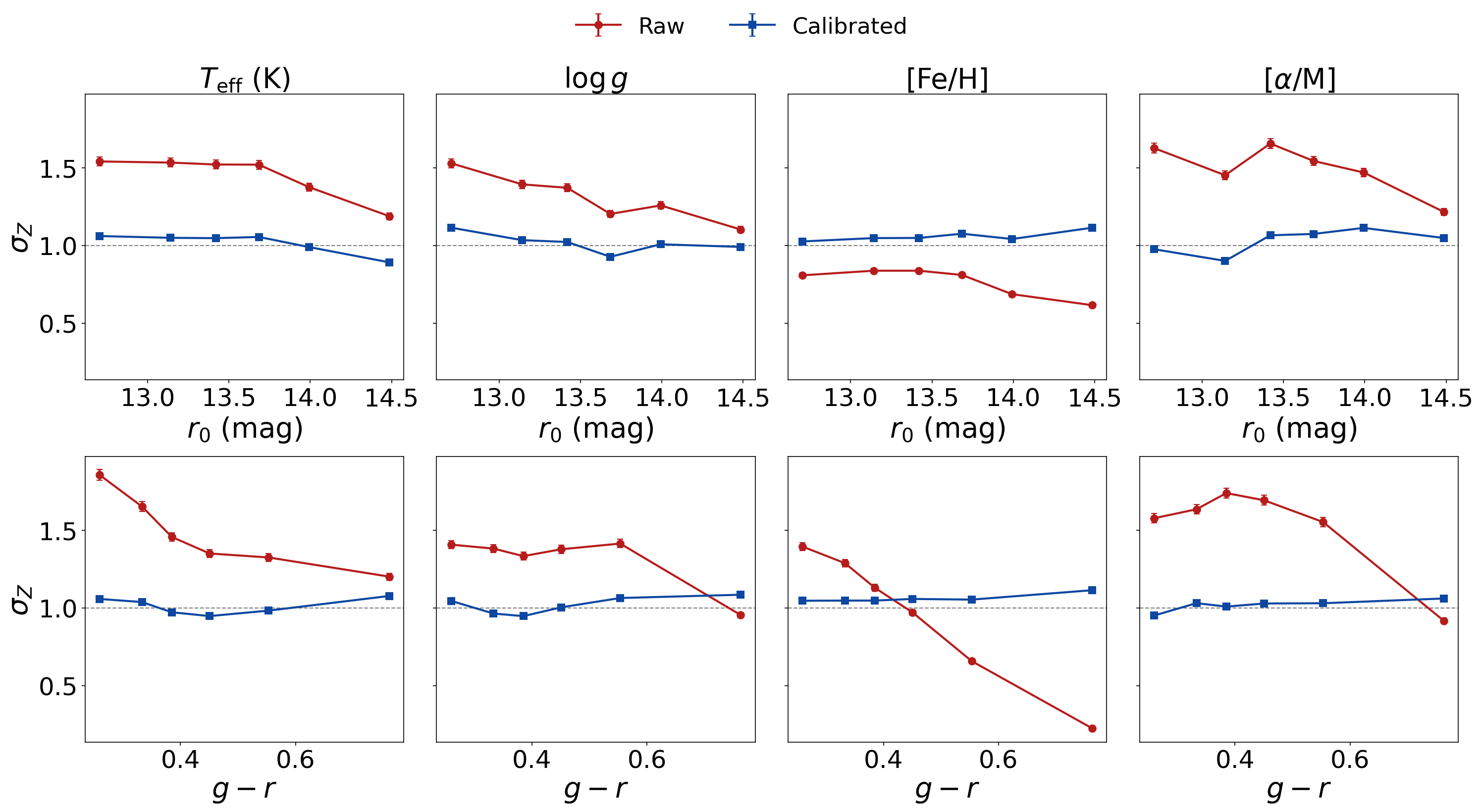}
		\caption{$\sigma_Z$ of the test sample as a function of the extinction-corrected $r$ magnitude (top) and of the $g-r$ colour (bottom), for the raw (red circles) and calibrated (blue squares) uncertainties of the colour models. The dashed line marks $\sigma_Z = 1$.}
		\label{fig:sigmaZ_mag}
	\end{figure*}

	The calibration of Sect.~\ref{sec:Factor_calibracion} depends only on the $g-r$ colour. Figure~\ref{fig:sigmaZ_mag} shows $\sigma_Z$ for the test sample in bins of $r_0$ and of $g-r$, for the raw and the calibrated uncertainties of the colour models. The calibrated $\sigma_Z$ stays close to 1 in all the colour bins, with a mild underestimation of the uncertainties for the reddest stars ($\sigma_Z \approx 1.06$--1.12 at $g-r \approx 0.75$~mag). It also stays close to 1 (0.89--1.12) over the magnitude range of the test sample: the uncertainties are slightly underestimated at the bright end for \logg\ and at the faint end for \feh, and slightly overestimated at the faint end for \teff. The raw $\sigma_Z$, in contrast, decreases with magnitude for the four parameters, down to 0.62--1.22 at $r_0 \approx 14.5$~mag, as expected if the intrinsic scatter of the relations, which the MC does not include, is independent of magnitude while the photometric errors grow with it.

	\subsection{Correlations between parameters}\label{sec:correlations}

\begin{figure}[!htbp]
		\centering
		\includegraphics[width=\columnwidth]{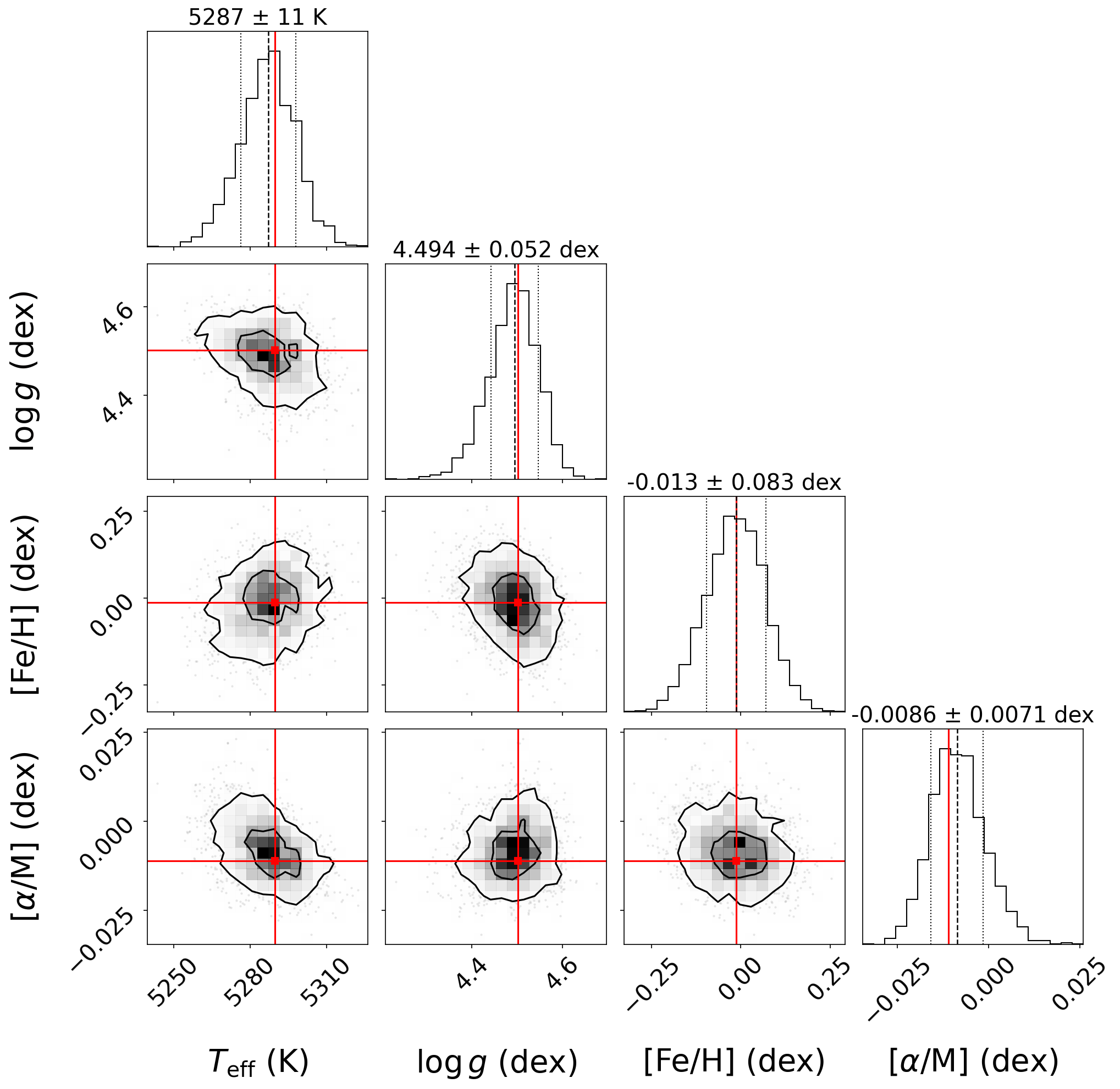}
		\caption{Corner plot of the MC realisations of the predicted stellar parameters (raw uncertainties) for the test-sample star \texttt{TILE\_ID} = 187221, \texttt{NUMBER} = 24521 ($r = 13.25$~mag). The contours in the two-dimensional panels enclose 39.3 and 86.5\,\% of the realisations; in the one-dimensional panels the dashed line marks the median and the dotted lines $\pm\smad$, whose values are given in the titles. The red lines and points mark the LAMOST values.}

		\label{fig:corner_plot}
	\end{figure}

	Figure~\ref{fig:corner_plot} illustrates the outcome of the MC simulation for one star of the test sample (\jplus\ \texttt{TILE\_ID} = 187221, \texttt{NUMBER} = 24521; $r = 13.25$~mag). The one-dimensional distributions are close to Gaussian, which justifies describing each of them with a single scale, and the LAMOST values of the four parameters lie within the 16th--84th percentile range of the MC distributions. The tilt of the two-dimensional distributions shows the correlations induced by the error propagation: for this star, \teff\ and \feh\ are positively correlated, whereas \feh\ and \am, \teff\ and \am, and \logg\ and \feh\ are anticorrelated.

\begin{figure}[!htbp]
		\centering
		\includegraphics[width=\columnwidth]{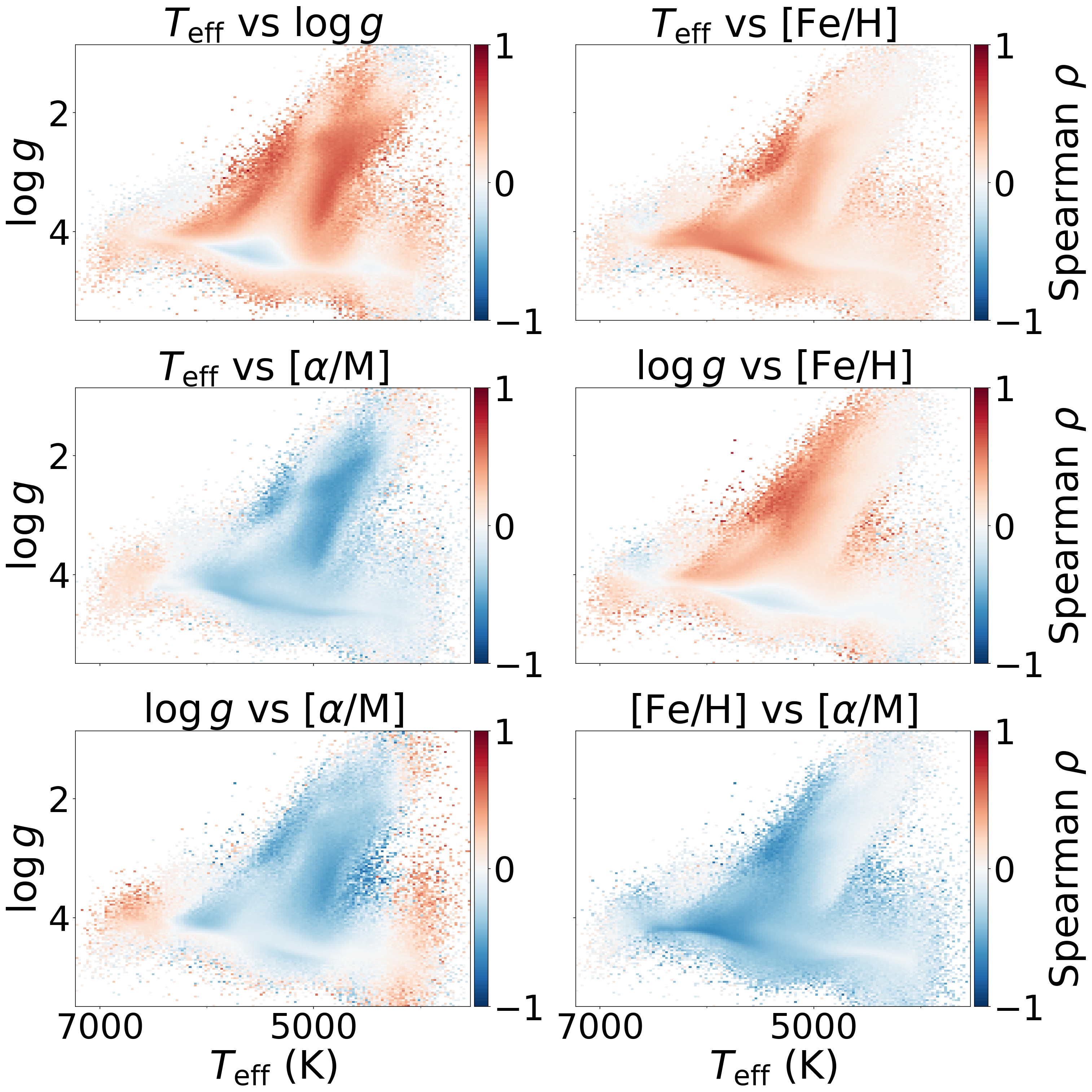}
		\caption{Median Spearman correlation induced by the error propagation between each pair of stellar parameters, in bins of the \teff--\logg\ plane, for the recommended sample of the colour models.}
		\label{fig:correlations}
	\end{figure}

	Figure~\ref{fig:correlations} maps the median Spearman correlation coefficient, $\rho$, between each pair of parameters across the Kiel diagram for the recommended sample of the colour models. The sign and the strength of the correlations depend on the location of the star in the diagram. \teff\ and \logg\ are positively correlated along the giant branch and the subgiant region, and anticorrelated along the main sequence. \teff\ and \feh\ are positively correlated almost everywhere, most strongly on the main sequence between 5000 and 6000~K and on the lower giant branch, and \teff\ and \am\ show the mirror pattern, an anticorrelation that is strongest on the upper giant branch. \logg\ and \feh\ are strongly correlated for the giants and weakly anticorrelated for the dwarfs, while \logg\ and \am\ are anticorrelated for the giants and nearly uncorrelated on the main sequence. Finally, \feh\ and \am\ are anticorrelated throughout the diagram, with $\rho \approx -0.5$ to $-0.75$ for the giants and the main sequence; this is the pair whose correlation most affects the chemical separation of Sect.~\ref{Componentes_MW}. The maps of the PC models (Fig.~\ref{fig:correlations_PCs}) are nearly identical, with slightly stronger correlations, which we attribute to the fact that every PC mixes all 66 colours, whereas each colour model uses a handful of explicit colours.

	\section{Properties of the recommended sample}\label{sec:recommended}

\begin{figure}[t]
    \centering
    \includegraphics[width=\columnwidth]{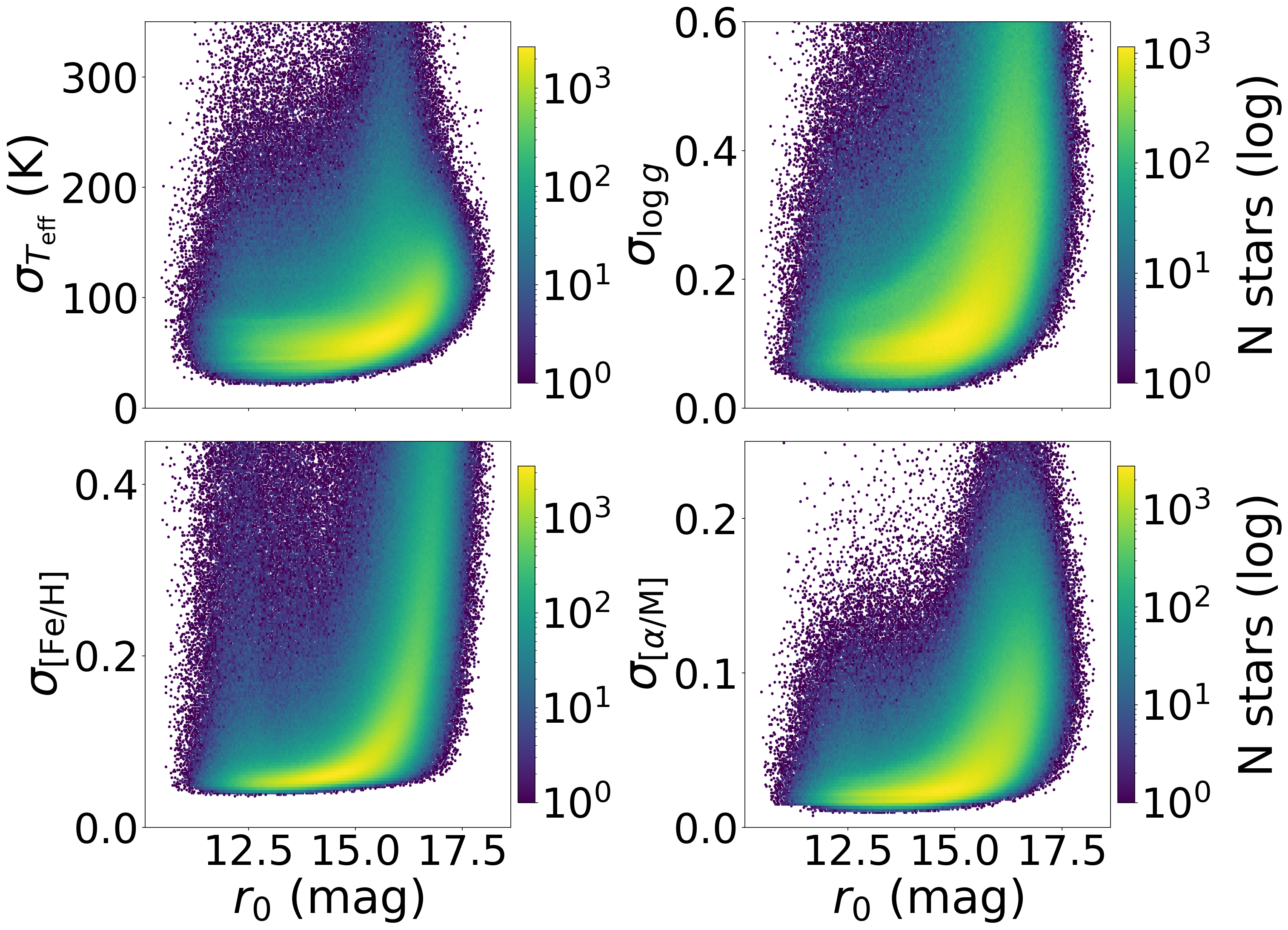}
    \caption{Calibrated uncertainties as a function of the extinction-corrected $r$-band magnitude for the recommended sample of the colour models (4,804,841 stars). The colour scale shows the number of stars per cell on a logarithmic scale; the upper limit of each panel is the quality cut of Sect.~\ref{sec:flags}.}
    \label{fig:sigma_vs_rcorr_todos}
\end{figure}

\begin{figure}[!htbp]
		\centering
		\includegraphics[width=\columnwidth]{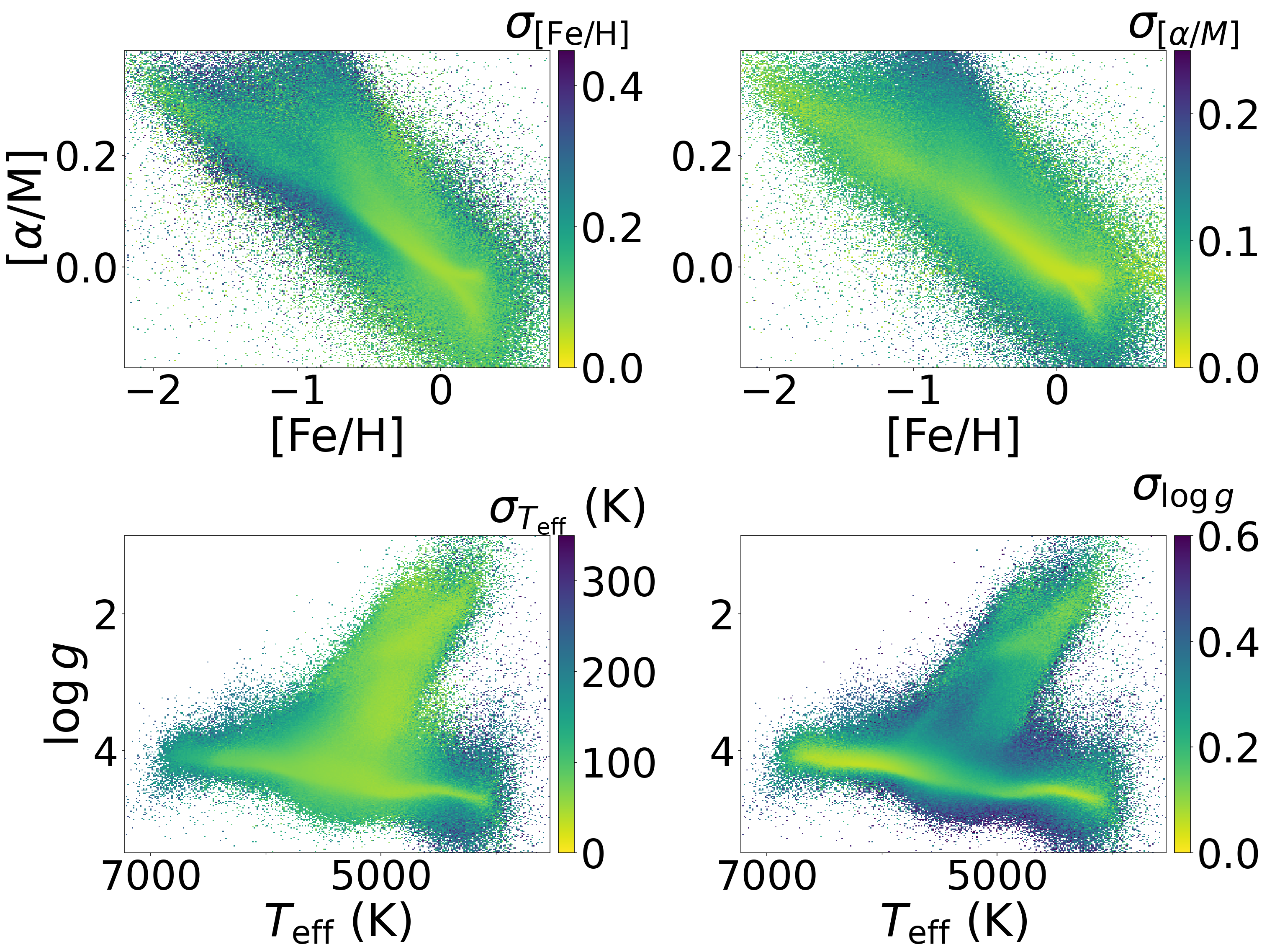}
		\caption{Median calibrated uncertainty per bin of the \feh--\am\ (top) and \teff--\logg\ (bottom) planes for the recommended sample of the colour models (4,804,841 stars): $\sigma_{\mathrm{[Fe/H]}}$ (top left), $\sigma_{[\alpha/\mathrm{M}]}$ (top right), $\sigma_{T_{\mathrm{eff}}}$ (bottom left), and $\sigma_{\log g}$ (bottom right).}
		\label{fig:sigma_bins}
	\end{figure}

Figure~\ref{fig:sigma_vs_rcorr_todos} shows the calibrated uncertainties of the recommended sample of the colour models as a function of the $r$-band magnitude, and Fig.~\ref{fig:sigma_bins} their distribution across the \feh--\am\ and \teff--\logg\ planes. The uncertainties are dominated by the photometric errors: they are smallest for the bright stars, where they approach the floor set by the calibration precision and the intrinsic scatter of the models, and they grow steeply beyond $r \approx 15$~mag, reaching the quality cuts at $r \approx 17.5$~mag. The recommended sample is therefore effectively limited to $r \lesssim 17.5$~mag, and its median uncertainties over most of that range are 70~K, 0.172~dex, 0.091~dex, and 0.038~dex in \teff, \logg, \feh, and \am, respectively. Except for \teff, these values are larger than those of the test sample (Sect.~\ref{sec:test}), which refer to stars with $G < 15$~mag. In the parameter planes, the uncertainties are smallest along the main sequence and the red clump, where the training sample is densest, and largest at the edges of the training domain: the hottest and coolest stars in \teff, the subgiant and upper-giant regions in \logg, and the metal-poor tail in \feh.

	\section{Principal-component models}\label{PCM}

\begin{figure*}[!tp]
		\centering
		\includegraphics[width=0.6\textwidth]{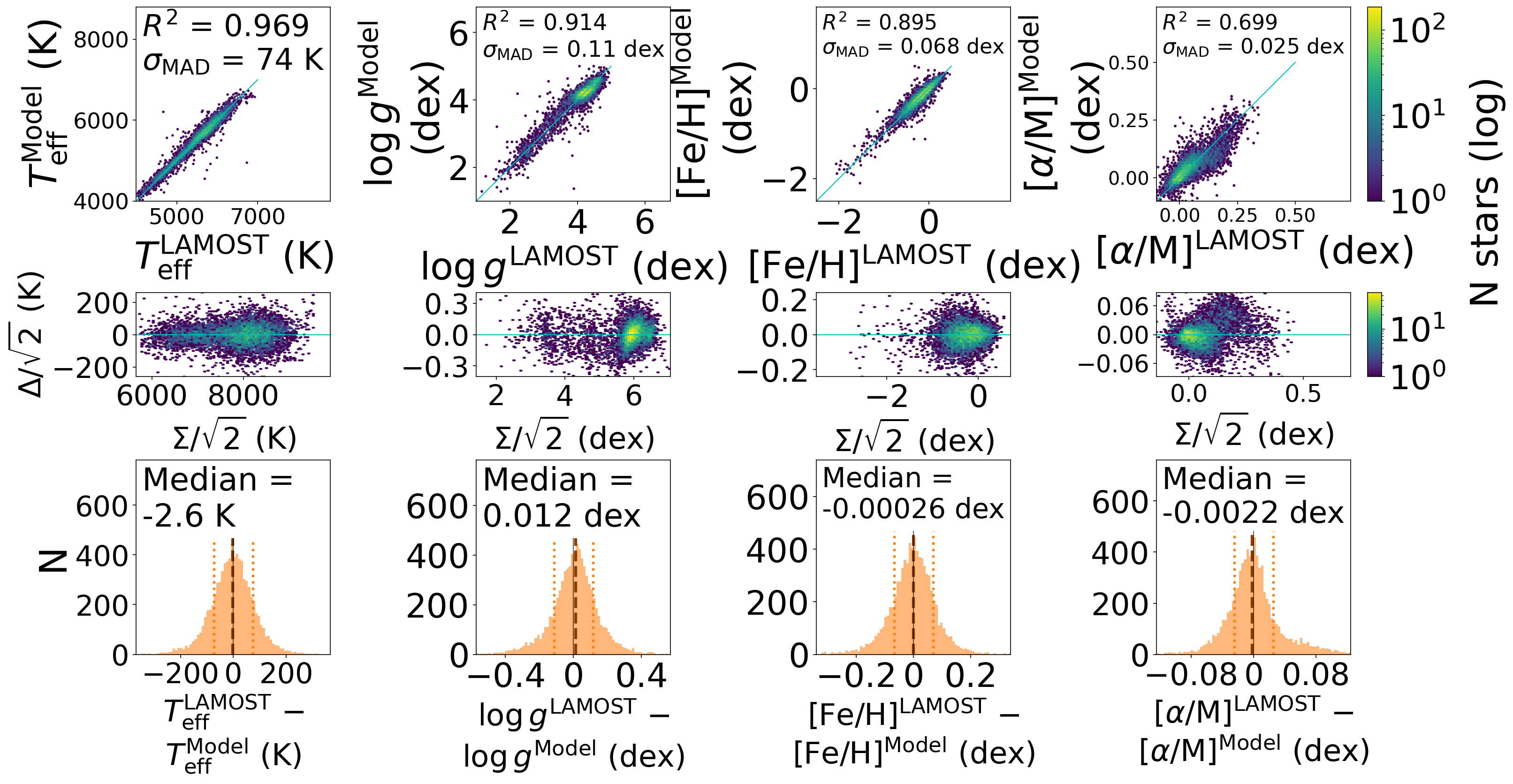}
		\caption{As Fig.~\ref{fig:BM}, for the PC models.}
		\label{fig:BM_PCs}

		\vspace{2mm}
		\includegraphics[width=0.6\textwidth]{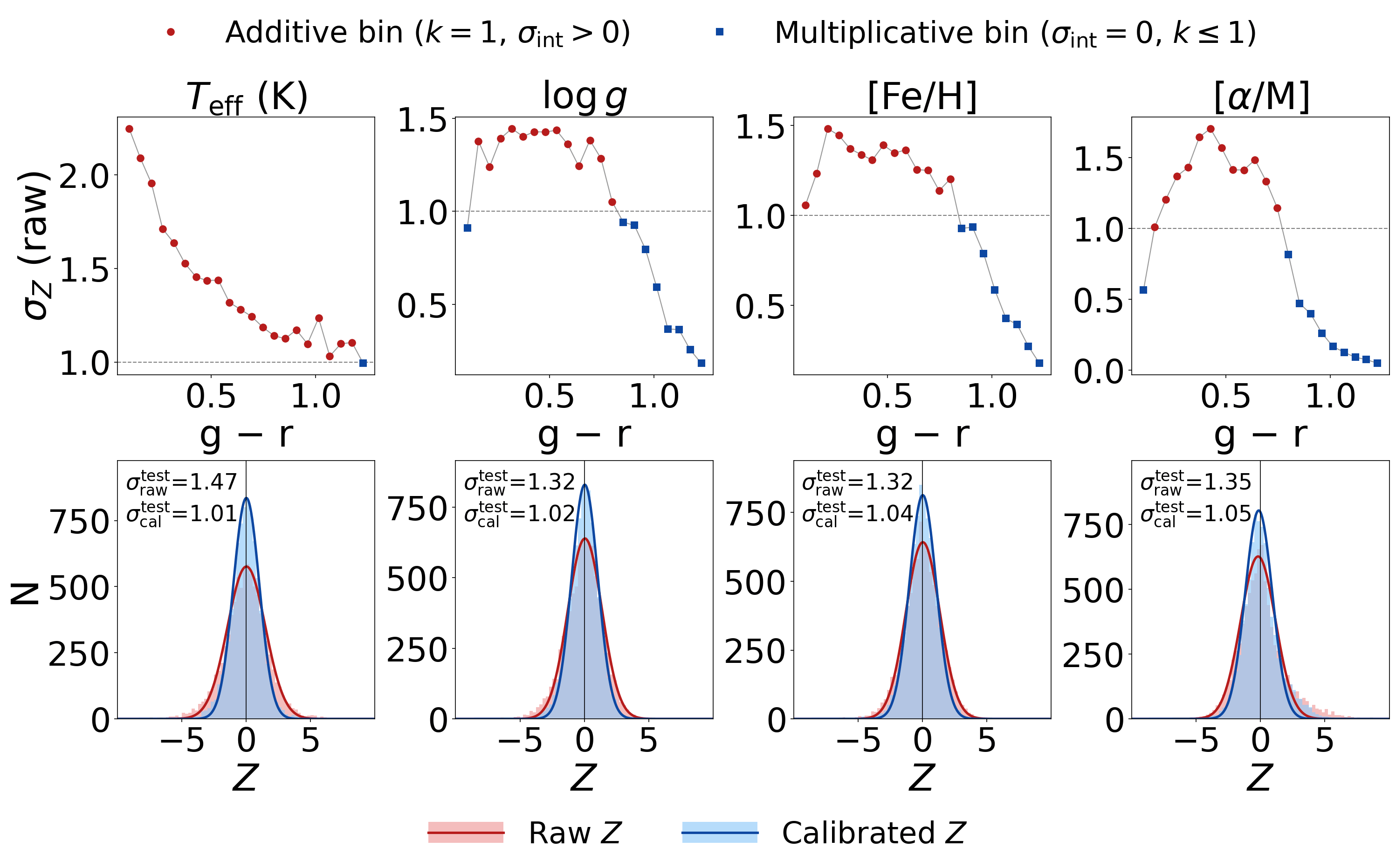}
		\caption{As Fig.~\ref{fig:Z}, for the PC models.}
		\label{fig:Z_PCs}
	\end{figure*}

\begin{figure*}[!tp]
		\begin{minipage}[t]{\columnwidth}
			\centering
			\includegraphics[width=\columnwidth]{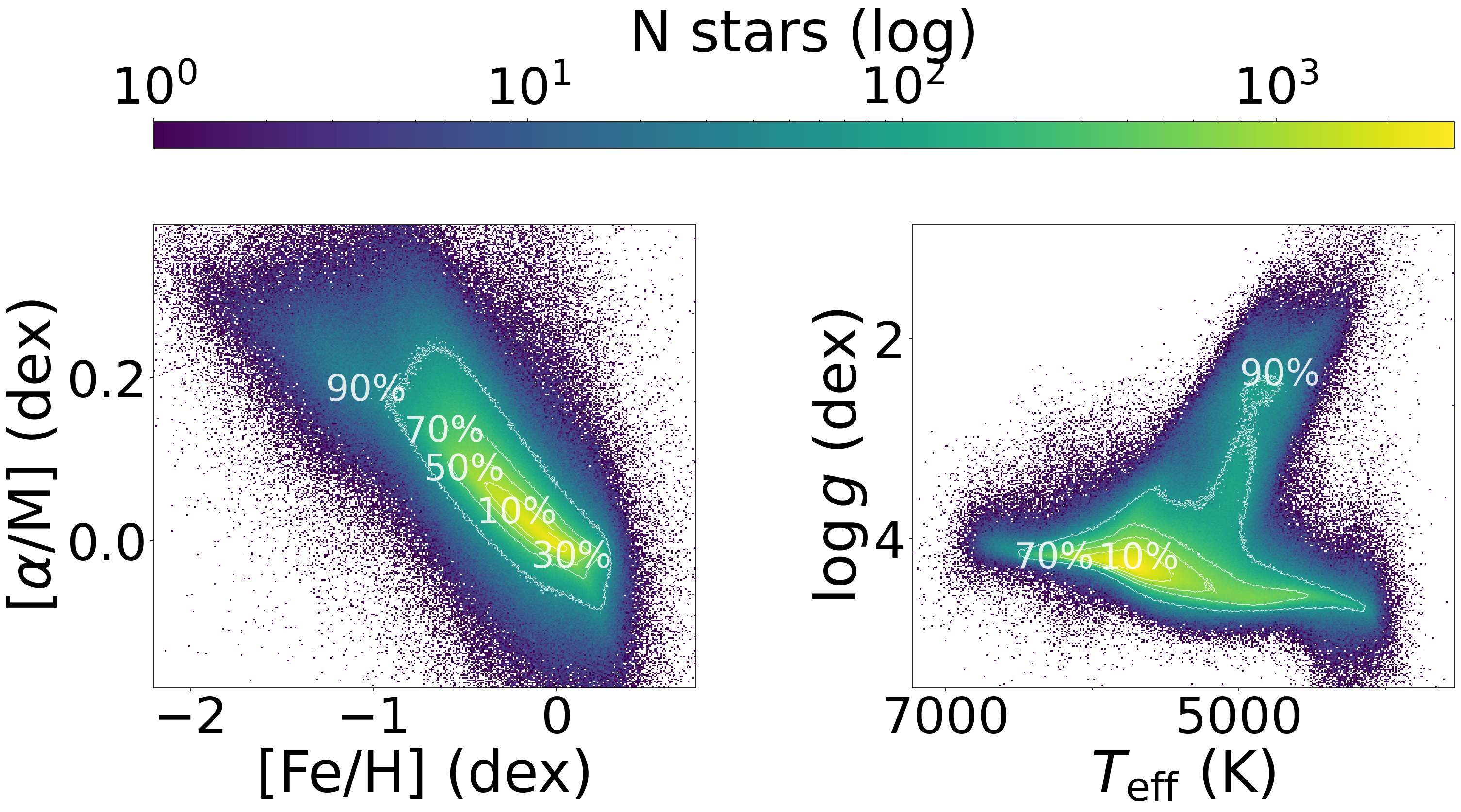}
			\caption{As Fig.~\ref{fig:Predictions}, for the recommended sample of the PC models (4,862,372 stars).}
			\label{fig:Predictions_PCs}
		\end{minipage}\hfill
		\begin{minipage}[t]{\columnwidth}
			\centering
			\includegraphics[width=\columnwidth]{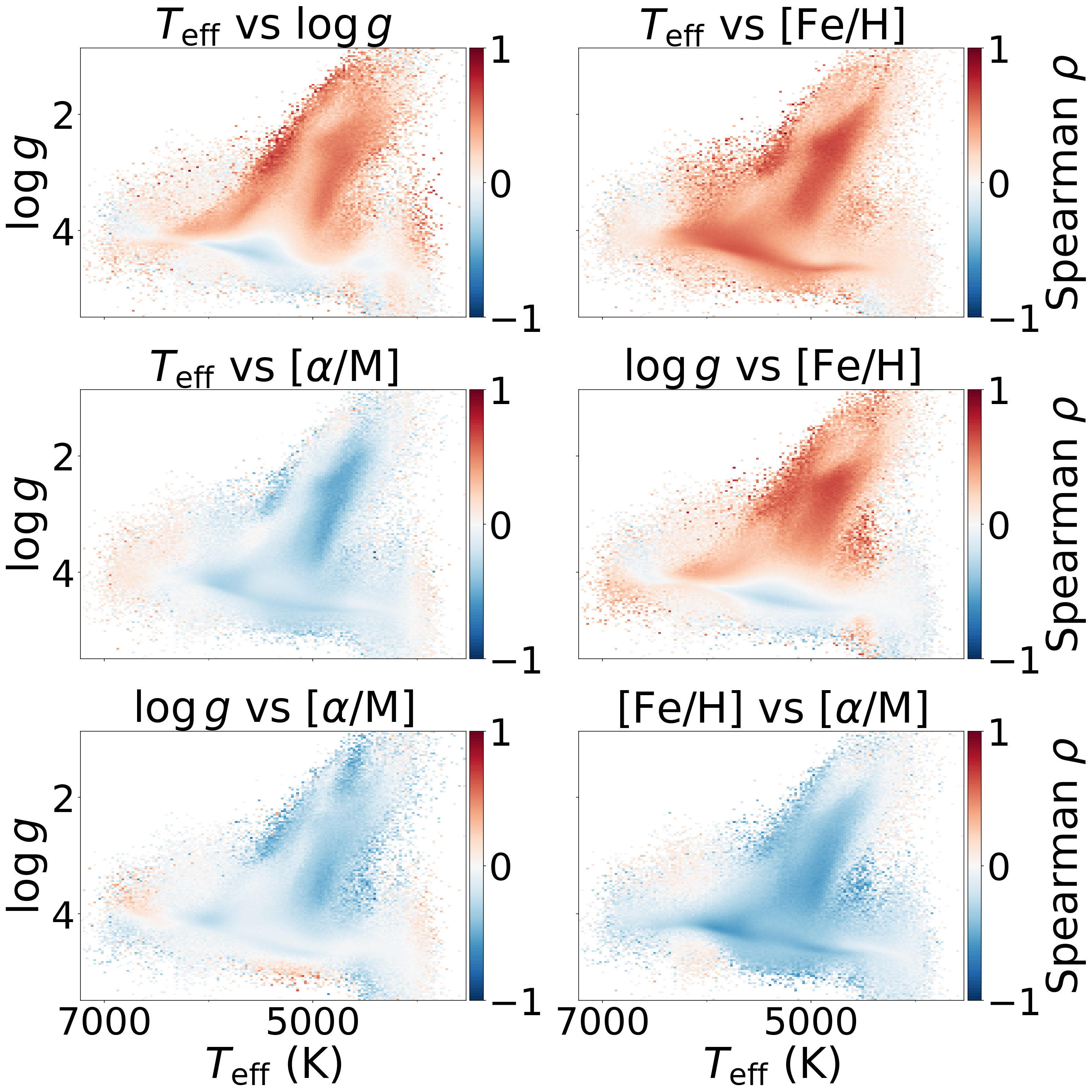}
			\caption{As Fig.~\ref{fig:correlations}, for the PC models.}
			\label{fig:correlations_PCs}
		\end{minipage}
	\end{figure*}

	This appendix presents, for the PC models, the figures and tables shown in the main text for the colour models: the adopted models (Table~\ref{tab:best_models_PCs}), the comparison with the LAMOST test sample (Fig.~\ref{fig:BM_PCs}), the calibration of the uncertainties (Fig.~\ref{fig:Z_PCs}), the quality flags (Table~\ref{tab:flags_PCs}), the distribution of the recommended sample in the parameter planes (Fig.~\ref{fig:Predictions_PCs}), and the induced correlations (Fig.~\ref{fig:correlations_PCs}).

	The four PC models use PC1 to PC5 and PC11, two or three of them also select PC7, PC8, or PC10, and only one selects PC6 or PC9, although PC1 alone accounts for 86\,\% of the variance of the colours and the components beyond PC7 for only 1.4\,\%. The information on the stellar parameters is therefore not concentrated in the components of largest variance, which is why we kept all 11 PCs as candidate variables.

	On the test sample, the PC models reach $R^2 = 0.969$, 0.914, 0.895, and 0.699 and $\smad = 74$~K, 0.11~dex, 0.068~dex, and 0.025~dex for \teff, \logg, \feh, and \am, the same values as the colour models to within the rounding, and they show the same features in every comparison: the zigzag pattern in \logg\ inherited from the training labels and the compression of the $\alpha$-enhanced population. Against APOGEE DR17 (13,450 stars), GALAH DR4 (1,525 stars), and \citet{2024ApJ...974..192H} (658,263 stars), they reach $R^2 = 0.972$, 0.940, 0.916, and 0.462; 0.953, 0.812, 0.834, and 0.082; and 0.967, 0.885, 0.876, and 0.546 for \teff, \logg, \feh, and \am, respectively, with offsets and \smad\ similar to those of the colour models (Table~\ref{tab:validation}).  The PC models exclude from the recommended sample the bright stars with non-zero photometric flags (mostly saturated pixels), for which the models extrapolate strongly, and these stars therefore drop out of the cross-matches with the surveys. The colour models are less affected because each of them uses only a subset of the colours, whereas every PC combines all of them. For the raw MC uncertainties we obtain $\sigma_{Z} = 1.47$, 1.32, 1.32, and 1.35, and for the calibrated ones $\sigma_{Z} = 1.01$, 1.02, 1.04, and 1.05 on the test sample. The recommended sample of the PC models contains 4,862,372 sources, 1.2\,\% more than that of the colour models.

	\begin{table*}[!tp]
		\centering
		\caption{As Table~\ref{tab:best_models}, for the PC models.}
		\label{tab:best_models_PCs}
		\scriptsize
		\setlength{\tabcolsep}{4pt}
		\begin{tabular}{lcccccl}
			\hline\hline
			Parameter & $g$ & $n$ & $N_{\rm coef}$ & $R^2$ & $Q$ & Variables \\
			\hline
			\teff & 4 & 7  & 330  & 0.9736 & 6104     & PC1, PC2, PC3, PC4, PC5, PC8, PC11 \\
			\logg & 4 & 8  & 495  & 0.9261 & 0.01913  & PC1, PC2, PC3, PC4, PC5, PC7, PC10, PC11 \\
			\feh  & 4 & 9  & 715  & 0.9183 & 0.006071 & PC1, PC2, PC3, PC4, PC5, PC6, PC7, PC8, PC11 \\
			\am   & 4 & 10 & 1001 & 0.7170 & 0.001835 & PC1, PC2, PC3, PC4, PC5, PC7, PC8, PC9, PC10, PC11 \\
			\hline
		\end{tabular}
	\end{table*}

	\begin{table*}[!tp]
		\caption{As Table~\ref{tab:flags}, for the PC models.}
		\label{tab:flags_PCs}
		\centering
		\small
		\setlength{\tabcolsep}{4pt}
		\begin{tabular}{l rr rr rr rr rr}
			\hline\hline
			\noalign{\smallskip}
			& \multicolumn{2}{c}{In validity range} & \multicolumn{2}{c}{flag 0} & \multicolumn{2}{c}{flag 1} & \multicolumn{2}{c}{flag 2} & \multicolumn{2}{c}{flag 3} \\
			Parameter & $N$ & \% & $N$ & \% & $N$ & \% & $N$ & \% & $N$ & \% \\
			\noalign{\smallskip}
			\hline
			\noalign{\smallskip}
			\teff   & 10,798,398 & 73.82 & 6,488,718 & 44.36 & 3,829,686 & 26.18 & 4,309,680 & 29.46 & 566     & 0.00 \\
			\logg   & 9,326,055  & 63.75 & 5,232,357 & 35.77 & 5,296,146 & 36.20 & 4,093,698 & 27.98 & 6,449   & 0.04 \\
			\feh    & 9,164,054  & 62.64 & 5,199,013 & 35.54 & 5,456,108 & 37.30 & 3,965,041 & 27.10 & 8,488   & 0.06 \\
			\am     & 7,754,386  & 53.01 & 5,327,588 & 36.42 & 6,752,856 & 46.16 & 2,426,798 & 16.59 & 121,408 & 0.83 \\
			\noalign{\smallskip}
			\hline
			\noalign{\smallskip}
			All four parameters & 7,353,149 & 50.27 & 4,862,372 & 33.24 & & & & & & \\
			\noalign{\smallskip}
			\hline
		\end{tabular}
	\end{table*}

\end{appendix}

\end{document}